\documentclass[preprint]{aastex}

\usepackage{lscape,natbib,amssymb}
\newcommand{\amm}{NH$_3$}
\newcommand{\dia}{N$_2$H$^+$}
\newcommand{\ddia}{N$_2$D$^+$}
\newcommand{\hd}{H$_2$D$^+$}

\newcommand{\kms}{km\,s$^{-1}$}
\newcommand{\cc}{cm$^{-3}$}
\newcommand{\vlsr}{$v_{\mbox{\tiny LSR}}$}
\newcommand{\one}{Paper I}
\newcommand{\two}{Paper II}
\newcommand{\showrev}{}

\shorttitle{Deuterium Fractionation in Oph B}
\shortauthors{Friesen et al.}

\begin{document}

\title{The Initial Conditions of Clustered Star Formation III. The Deuterium Fractionation of the Ophiuchus B2 Core}
\author{R. K. Friesen\altaffilmark{1,2,3}}
\altaffiltext{1}{National Radio Astronomy Observatory, 520 Edgemont Rd, Charlottesville VA 22903}
\altaffiltext{2}{Department of Physics and Astronomy, University of Victoria, PO Box 3055, STN CSC, Victoria BC CANADA V8W 3P6}
\email{rfriesen@nrao.edu}

\author{J. Di Francesco\altaffilmark{3,2}}
\altaffiltext{3}{National Research Council Canada, Herzberg Institute of Astrophysics, 5071 West Saanich Road, Victoria, British Columbia, Canada V9E 2E7}

\author{P. C. Myers\altaffilmark{4}}
\altaffiltext{4}{Harvard-Smithsonian Center for Astrophysics, 60 Garden Street, Cambridge, MA 02138}

\author{A. Belloche\altaffilmark{5}}
\altaffiltext{5}{Max-Planck Institut f{\"u}r Radioastronomie, Auf dem H{\"u}gel 69, 53121 Bonn, Germany}

\author{Y. L. Shirley\altaffilmark{6}}
\altaffiltext{6}{Steward Observatory, University of Arizona, 933 N. Cherry Ave., Tucson, AZ 85721}

\author{T. L. Bourke\altaffilmark{4}}

\author{P. Andr\'{e}\altaffilmark{7}}
\altaffiltext{7}{Laboratoire AIM, CEA/DSM-CNRS-Universit{\'e} Paris Diderot, IRFU/Service d'Astrophysique, C.E. Saclay, Orme des Merisiers, 91191 Gif-sur-Yvette, France}

\begin{abstract}

We present \ddia\, 3-2 (IRAM) and \hd\, $1_{11} - 1_{10}$ and \dia\, 4-3 (JCMT) maps of the small cluster-forming Ophiuchus B2 Core in the nearby Ophiuchus molecular cloud. In conjunction with previously published \dia\, 1-0 observations, the \ddia\, data reveal the deuterium fractionation in the high density gas across Oph B2. The average deuterium fractionation $R_D = N(\mbox{\ddia})/N(\mbox{\dia}) \sim 0.03$ over Oph B2, with several small scale $R_D$ peaks and a maximum $R_D = 0.1$.  The mean $R_D$ is consistent with previous results in isolated starless and protostellar cores. The column density distributions of both \hd\, and \ddia\, show no correlation with total H$_2$ column density. We find, however, an anticorrelation in deuterium fractionation with proximity to the embedded protostars in Oph B2 to distances $\gtrsim 0.04$\,pc. Destruction mechanisms for deuterated molecules require gas temperatures greater than those previously determined through \amm\, observations of Oph B2 to proceed. We present temperatures calculated for the dense core gas through the equating of non-thermal line widths for molecules (i.e., \ddia\, and \hd) expected to trace the same core regions, but the observed complex line structures in B2 preclude finding a reasonable result in many locations. This method may, however, work well in isolated cores with less complicated velocity structures. Finally, we use $R_D$ and the \hd\, column density across Oph B2 to set a lower limit on the ionization fraction across the core, finding a mean $x_{e, lim} \gtrsim few \times 10^{-8}$. Our results show that care must be taken when using deuterated species as a probe of the physical conditions of dense gas in star-forming regions. 


\end{abstract}


\section{Introduction}

The Ophiuchus molecular cloud is a source of ongoing low-mass clustered star formation relatively close to the Sun \citep[$d \sim 120$\,pc;][]{lombardi08}. Through low resolution DCO$^+$ observations, \citet{loren90} discovered a number of dense Cores (labelled A through F) in the central Ophiuchus region with masses of a few to several tens of solar masses, which represented a range in evolutionary status given the presence or absence of embedded protostars\footnotemark\footnotetext{In this paper, we describe Oph A, B, C, etc., as `Cores' since this is how these features were named in DCO$^+$ observations of the L1688 region by \citet{loren90}. Since then, higher-resolution data such as those described in this paper have revealed substructure in these features that could be themselves precursors to stars, i.e., cores.  To avoid confusion, we refer the larger features in Oph as `Cores' and Core substructure identified by {\sc clumpfind} as `clumps'.}. Large-scale submillimetre and millimetre observations of Ophiuchus identified several new Cores, and revealed the Cores were further fragmented into multiple dense clumps, each with masses $M \lesssim 1$\,M$_\odot$ \citep{motte98,johnstone00,johnstone04,young06,sadavoy10}. Analysis of the clump kinematics suggests most are in virial equilibrium and hence not transient objects \citep{andre07}. Recent Spitzer Space Telescope studies of infrared emission towards the Cores have characterized in detail the starless and protostellar clump population \citep{jorgensen08,enoch09}. These objects are thus ideal locations to study the physical characteristics of dense cores in the process of forming small stellar clusters. 

To characterize the physical properties of the dense, cluster forming Ophiuchus Cores, we have previously studied Oph B, C and F through high resolution \amm\, (1,1) and (2,2) observations \citep[][hereafter \one]{friesen09nh3}, and Oph B in both single-dish and interferometer \dia\, 1-0 emission \citep[][hereafter \two]{friesen10}. The \amm\, and \dia\, observations probe gas at densities $n \sim 10^{3-4}$\,\cc\, and $n \sim 10^5$\,\cc, respectively. 

In \one, we identified \citep[through the 3D {\sc clumpfind} structure-finding algorithm; ][]{williams94} a number of individual clumps in \amm\, emission in the Cores, which did not correlate well with those identified in continuum emission. We determined that Oph B and F were warmer on average than typically found for isolated cores (mean $T_K \sim 15$\,K), and we found no trend with temperatures and H$_2$ column density calculated from submillimetre continuum emission. In contrast, Oph C is colder, with temperatures decreasing to a minimum of  9\,K at the dust continuum emission peak. Line widths in both the Oph B1 and B2 sub-Cores were found to be wide and transonic, despite the presence of three embedded protostars in B2 and none in B1. Oph F, with four embedded protostars, has regions of small and large line widths and complicated line structures. Again similar to isolated objects, \amm\, line widths in Oph C are extremely narrow and consistent with being due to purely thermal motions. In an interesting result, the fractional abundance of \amm\, with respect to H$_2$, $X(\mbox{\amm})$, appeared to decrease with increasing H$_2$ column density, $N(\mbox{H$_2$})$, suggesting \amm\, may be depleted towards locations of high column density and thus may not be tracing the densest core gas. Finally, \one\, revealed that cores with characteristics of both isolated and clustered environments can coexist in a clustered star forming environment.

In \two, we again used {\sc clumpfind} to identify small-scale structure in \dia\, emission from Oph B and found that the \dia\, clumps matched well \amm\, clumps from \one, and were consequently again offset from continuum objects. In both studies, little difference was found in the gas properties towards embedded protostars compared with the general core gas. Line widths in Oph B2 remain large at the densities traced by \dia. In B1, however, \dia\, line widths are significantly narrower than found in \amm\, emission. Applying the gas temperatures determined in \one, we found that motions in Oph B1 were subsonic with a mean ratio of non-thermal line width to sound speed $\sigma_{\mbox{\tiny NT}} / c_s = 0.7$ across the core, while in B2 $\sigma_{\mbox{\tiny NT}} / c_s = 1.3$. A clear trend of decreasing $X(\mbox{\dia})$ with increasing $N(\mbox{H$_2$})$ was found towards Oph B2, with emission suggestive of a \dia\, hole towards the continuum emission peak found in high resolution \dia\, 1-0 emission observed with the Australia Telescope Compact Array. 

The results of \amm\, and \dia\, observations towards Oph B2 in \one\, and \two\, indicate that the highest density gas may best be probed by molecular emission lines excited at higher critical densities, or by deuterated molecules which are expected to be abundant in cold cores where significant depletion has occurred. In cold molecular cloud cores, the deuteration process is initiated by the reaction

\begin{equation}
\mbox{H$_3^+$} + \mbox{HD} \rightleftharpoons \mbox{\hd} + \mbox{H$_2$} + \Delta E
\label{eqn:hd}
\end{equation}

\noindent which is exothermic in the forward direction with $\Delta E = 232$\,K \citep{millar89}. At very low temperatures ($\lesssim 20$\,K), the forward reaction dominates, making \hd\, a key molecular ion in the enhancement of deuterated species in a molecular core. \showrev{CO, which can destroy both \hd\, and H$_3^+$ via proton-transfer reactions, is additionally expected to deplete from the gas phase through freeze-out onto dust grains. The deuterium fractionation can thus be elevated orders of magnitude above the cosmic [D]/[H] ratio of $10^{-5}$ for H$_3^+$, as well as for other species which react with its deuterated forms. }

\showrev{The ortho-\hd\, $1_{11}-1_{10}$ transition at 372\,GHz was first detected towards the low mass young stellar object (YSO) NGC 1333 IRAS 4A by \citet{stark99}. Multiple detections of \hd\, and doubly-deuterated H$_3^+$, D$_2$H$^+$, have followed in both starless and protostellar cores \citep{caselli03,stark04,vastel04}, including a recent survey of \hd\, emission towards sixteen objects by  \citet{caselli08}.  \citet{walmsley04} show the \hd\, abundance determined for the dark core L1544 by \citealt{caselli03} is consistent with a `complete depletion' model, where heavy elements such as C, N and O (and, consequently, molecules containing these elements) are missing from the gas phase at the highest core densities. In this scenario where no heavy elements are left in the gas phase, \hd\, and multiply-deuterated forms of H$_3^+$ are the only gas tracers of the physical conditions of the dense core.}

\showrev{Based on our previous work in \one\, and \two, Oph B2 is a good target to investigate the distribution of deuterated species in a more complex star-forming core. }To this end, we present in this paper \ddia\, 3-2, \dia\, 4-3 and \hd\, $1_{11} - 1_{10}$ emission maps of Oph B2. In the following sections, we present the observations in \S\ref{sec:obs}, and discuss the observed distributions of \ddia\, 3-2, \hd\, $1_{11} - 1_{10}$ and \dia\, 4-3 in \S\ref{sec:results}. We fit the observed spectra with a single Gaussian (for \hd) or a multiple-component Gaussian (for \ddia\, and \dia\, due to their hyperfine line structure) and discuss the line velocity centroids and widths in \S\ref{sec:analysis_d}, and also present column density and fractional abundance calculations. In \S\ref{sec:discussion} we look at the non-thermal line widths in Oph B2 as a function of critical density of the observed emission lines, and discuss trends in the deuterium fractionation. Finally, we present a lower limit to the electron abundance in B2, and summarize our results in \S\ref{sec:summary}. 


\section{Observations}
\label{sec:obs}

In the following section we discuss the observations presented in this paper. We list the species, transitions, and rest frequencies observed in Table \ref{tab:obs}. 

\subsection{\ddia\, at IRAM}

\label{sec:n2d_obs}

An On-The-Fly (OTF) map of the \ddia\, 3-2 line at 231.321\,GHz in Oph B2 was made at the Institut de Radio Astronomie Millim{\'e}trique (IRAM) 30\,m Telescope during the 2007 and 2008 winter semesters using the HEterodyne Receiver Array \citep[HERA, ][]{schuster04}. HERA is a multi-beam spectral line receiver with nine dual polarization pixels arranged in a $3 \times 3$ array, with pixels separated by 24\arcsec\, [approximately twice the 231\,GHz 11\arcsec\, beam full width half power (FWHP)]. The maps were performed with the array tracking the sky rotation, and rotated by 9.5\arcdeg\, with respect to the equatorial system, such that the spacing between OTF lines was 4\arcsec. Observations were frequency-switched in-band with a throw of $2 \times 6.9$\,MHz. The Versatile Spectrometer Assembly (VESPA) autocorrelator was used as the backend, with a 20\,MHz bandwidth and 40\,kHz channel spacing, corresponding to 0.05\,\kms\, per channel at 231\,GHz. 

The data were calibrated to $T_A^*$ units at the telescope using the Multichannel Imaging and Calibration Software for Receiver Arrays (MIRA). Further calibration, including baselining and folding of the frequency-switched spectra, was performed using the CLASS software package\footnote{see http://www.iram.fr/IRAMFR/GILDAS.}. The data were smoothed by two spectral channels along the frequency axis to improve sensitivity, for a final spectral resolution of 80\,kHz, or 0.1\,\kms. A second-order polynomial baseline was fitted to the non-line channels and subtracted from the individual spectra before folding the data. The forward ($F_{eff}$) and beam ($B_{eff}$) efficiencies of the telescope were interpolated from determined values at 210\,GHz and 260\,GHz, giving $F_{eff} = 0.92$ and $B_{eff} = 0.58$ at 231\,GHz. The data were then calibrated to main beam temperature via the relation $T_{MB} = (F_{eff}\,/\,B_{eff})\, T_A^*$. For comparison with the \dia\, 1-0 results in \two, we further converted the data to units of Jy\,/\,beam$^{-1}$ and convolved the map to a final angular resolution of 18\arcsec\, to match the \dia\, data. 

As the observations were performed in a pooled mode, the 4\arcmin\, $\times$ 3\arcmin\, map was observed in three 4\arcmin\, $\times$ 1\arcmin\, strips to ensure good sensitivity in potentially limited time. Two-thirds of the map was completed in this manner in winter 2007, while the remaining strip was observed in winter 2008. The sensitivity of the map is consequently not entirely uniform, with the rms noise in the most southern 4\arcmin\, $\times$ 1\arcmin\, strip greater [0.15\,K ($T_{MB}$)] than in the rest of the map [0.08\,K ($T_{MB}$)]. Most emission was found in the central and northern map sections, however, so this noise increase does not affect our results significantly. 

\subsection{\hd\, and \dia\, at the JCMT}

Simultaneous observations of the \hd\, 1$_{10}-1_{11}$ and \dia\, 4-3 emission lines at 372.421\,GHz and 372.673\,GHz, respectively, were performed at the James Clerk Maxwell Telescope (JCMT) over the 07A - 08B semesters using the 16-receptor Heterodyne Array Receiver Program B-band receiver \citep[HARP-B, ][]{smith08}. HARP-B is a $4 \times 4$ pixel array, with array spacing of 30\arcsec\, and a 2\arcmin\, $\times$ 2\arcmin\, footprint. The JCMT beam at 372\,GHz is $\sim 13$\arcsec\, FWHM. We used the ACSIS correlator as the backend, configured to have a 400\,MHz effective bandwidth (e.g., wide enough to encompass both lines), tuned halfway in frequency between the two lines, and a frequency resolution of 61\,kHz, or 0.05\,\kms\, at 372.4\,GHz. Observations of five separate HARP footprints were performed in position-switching mode to create an undersampled map of \hd\, and \dia\, over Oph B2. The footprints were placed such that the final beam spacing on the sky is 21\arcsec\, with a coverage of $\sim 4$\arcmin\, $\times$ 2\arcmin. The fifth footprint provided additional coverage towards the location of peak continuum emission in B2, but has lower sensitivity compared with the other footprints. The observations were performed in band 1 weather only ($\tau_{225\,GHz} < 0.05$) and generally while Ophiuchus was above 30\arcdeg\, in elevation. The mean rms across the \hd\, map is 0.04\,K ($T_A^*$) and 0.05\,K ($T_A^*$) across the \dia\, map. The main beam efficiency factors, $\eta_{MB}$, of the HARP receptors at 372\,GHz are not currently well-known. At 345\,GHz, the mean $\eta_{MB} = 0.60$, with an rms variation of $\sim 5$\,\% between detectors \citep{buckle09}. We therefore estimate $\eta_{MB} \sim 0.6$ with an uncertainty of $\sim 10$\%.

Data reduction was performed using the Starlink software package. Each night's data were first checked for noisy or malfunctioning pixels, and these pixels were flagged individually for each integration. A linear baseline was then removed from each integration, and the data were combined into a final cube with 15\arcsec\, pixels. To increase sensitivity, the data were averaged along the velocity axis by 2 channels, giving a final velocity resolution of 0.1\,\kms. 

\section{Results}
\label{sec:results}

We first discuss the integrated intensity of \ddia\, 3-2, \hd\, $1_{11} - 1_{10}$ and \dia\, 4-3 over Oph B2 and compare the species distributions with submillimetre continuum emission and the locations of embedded protostars. 

\begin{figure}
\plotone{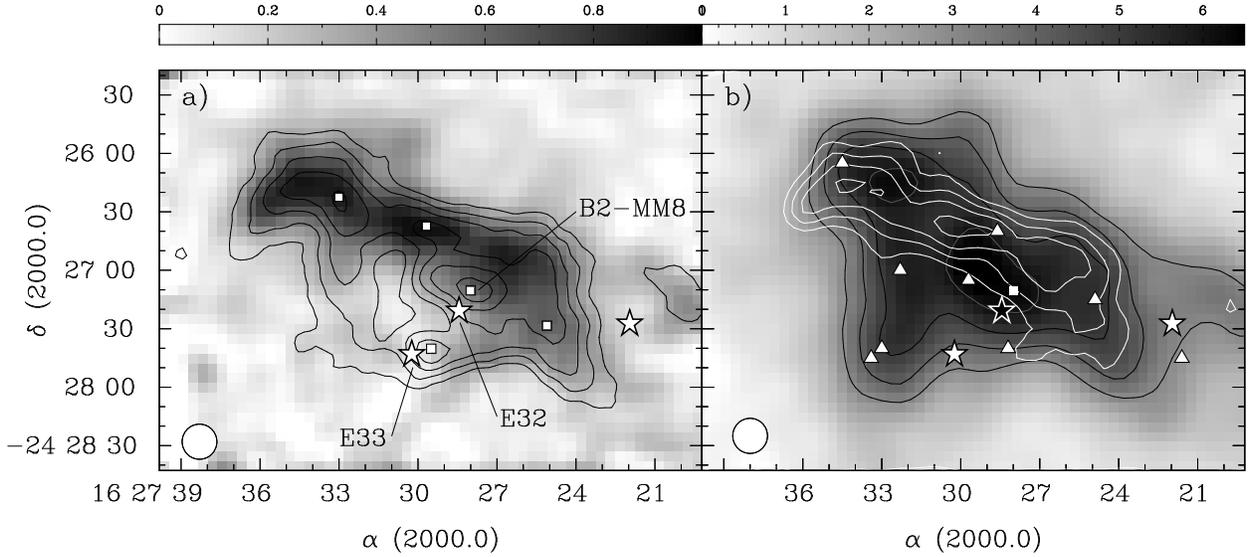} 
\caption[Integrated \ddia\, 3-2 emission and \dia\, 1-0 emission towards Oph B2]{a) Integrated \ddia\, 3-2 emission towards Oph B2 (greyscale) in $T_B$ units. Data have been convolved to a final 18\arcsec\, beam (FWHM). Black contours show 850\,\micron\, continuum emission in increments of 0.1\,Jy\,beam$^{-1}$. Squares show the locations of continuum clumps identified by \citet{jorgensen08}. The continuum clump B2-MM8 is identified. In both plots, stars indicate the positions of Class I protostars (with Elias 32 and 33 identified) and the 18\arcsec\, beam is shown at lower left. b) Integrated \dia\, 1-0 emission towards Oph B2 observed with the Nobeyama 45\,m Telescope at 18\arcsec\, (FWHM) resolution (greyscale). Black and grey contours begin at 3\,K\,\kms\, (T$_{MB}$) and increase by 1\,K\,\kms. Triangles show the peak locations of \dia\, clumps identified with {\sc clumpfind}, and the continuum clump B2-MM8 is shown by the square. White contours trace \ddia\, emission from a), beginning at 0.45\,K\,\kms\, ($T_B$) and increasing by 0.15\,K\,\kms. }
\label{fig:n2d}
\end{figure}

In Figure \ref{fig:n2d}a, we show the integrated \ddia\, 3-2 intensity (convolved to 18\arcsec\, resolution as described in \S\ref{sec:n2d_obs}) in Oph B2 against contours of 850\,\micron\, continuum emission observed at the JCMT at 15\arcsec\, FWHM resolution \citep[original map by \citealt{johnstone00}, combined with all other SCUBA archive data by][following the description in \citealt{kirk06}]{jorgensen08}.  Also identified are the locations of 850\,\micron\, continuum clumps \citep{jorgensen08} and Class I protostars [classifications and locations taken from \citealt{enoch09}, two of which were previously identified as Elias 32 and 33 (E32 and E33) by \citealt{elias78}; also VSSG 17 and 18 by \citealt{vssg}]. The \ddia\, emission is confined to the northern edge of B2, and generally follows the continuum contours in this region, but a very close correspondence is only seen towards the north-east tip of B2. Three maxima of integrated intensity are present, of which two are co-located with continuum clumps, while the third is seen to the north-west of the continuum peak. The \ddia\, emission is weak towards the continuum emission peak, and avoids entirely areas near the protostars.



In Figure \ref{fig:n2d}b, we plot the \ddia\, 3-2 integrated intensity contours over a map of \dia\, 1-0 integrated intensity, observed with the Nobeyama 45\,m Telescope at 18\arcsec\, resolution. The \dia\, observations and analysis were presented in \two. We also show the locations of \dia\, clumps, identified through {\sc clumpfind} in \two. Overall, \ddia\, emission is found within the extent of \dia\, 1-0 emission. In a similar fashion to the continuum emission, however, we find offsets between the locations of peak \ddia\, and \dia\, integrated intensity. In particular, the \ddia\, integrated intensity maximum is offset from the integrated \dia\, 1-0 maximum in B2 by $\sim 20$\arcsec\, to the northeast, $\sim 1$ beam width. The \dia\, 1-0 integrated intensity maximum is itself offset to the east from the continuum peak by $\sim 22$\arcsec\, (\two). The northeastern \ddia\, integrated intensity peak, which corresponds well with continuum contours, is also offset from the integrated \dia\, intensity peak by $\sim 20$\arcsec\, to the east. 

\begin{figure}
\plotone{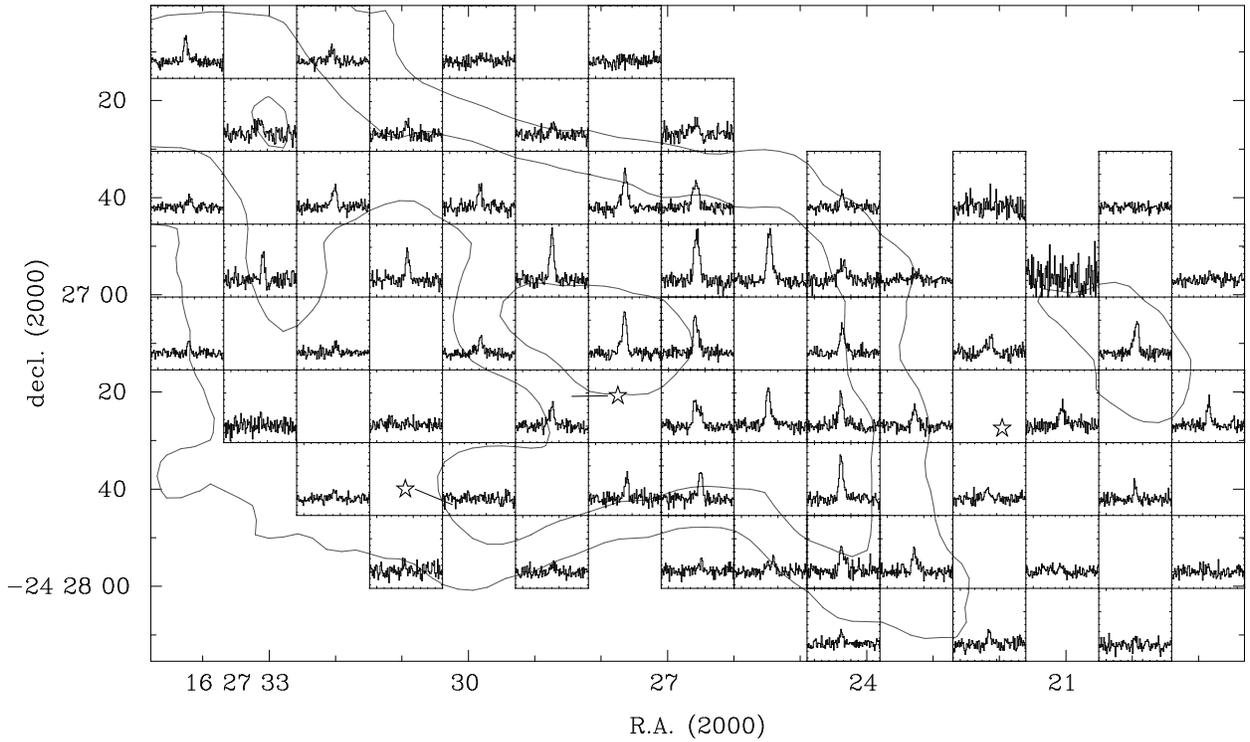} 
\caption{\hd\, $1_{11} - 1_{10}$ spectra observed with HARP at the JCMT towards Oph B2 with 15\arcsec\, FWHM. The grey contours show 850\,\micron\, continuum emission observed at the same resolution at 0.1\,Jy\,beam$^{-1}$, 0.3\,Jy\,beam$^{-1}$ and 0.5\,Jy\,beam$^{-1}$. Stars represent Class I protostars, with solid lines indicating the location if within a spectrum box. The velocity scale on the subplots runs from -1\,\kms to 9\,\kms, while the amplitude ranges from -0.25\,K to 0.85\,K ($T_{MB}$, assuming $\eta_{MB} = 0.6$). }
\label{fig:h2d}
\end{figure}

We show in Figure \ref{fig:h2d} \hd\, $1_{11} - 1_{10}$ spectra across Oph B2 with overlaid 850\,\micron\, continuum contours. We find that significant \hd\, emission is present over much of Oph B2, with a larger extent than found in \hd\, observations towards any other cores so far observed \citep{caselli03,vastel06,pagani09}. The \hd\, emission again follows generally the continuum emission but avoids the embedded protostars in B2, similar to \ddia. While relatively strong emission is seen towards the continuum peak, the \hd\, integrated intensity maximum (not shown) is also offset the continuum peak by an estimated $\sim 20-30$\arcsec\, (due to the map's undersampling we are unable to determine exactly the location of maximum integrated intensity). In contrast to the \ddia\, emission, the \hd\, integrated intensity maximum is found to the northwest rather than the northeast  (although \ddia\, emission is also found towards the \hd\, peak). The pixel-to-pixel variations seen in Oph B2 may not be significant, however, due to the $\sim 5$\,\% variation determined for the main beam efficiency factors between different HARP receptors \citep{buckle09}. We detect some \hd\, emission in north-east B2 where was found strong \ddia\, emission, but do not see a significant integrated intensity maximum. (Note, however, that the HARP footprints do not extend past an R.A. of 16:27:35, and are therefore not sensitive to any emission at the tip of the continuum contours in this region.) Overall, where \ddia\, emission appears strongest in the north and east of Oph B2, \hd\, emission is strongest towards B2's western edge. 

In Figure \ref{fig:n2h}, we show \dia\, 4-3 spectra across Oph B2 with overlaid 850\,\micron\, continuum emission contours. We find bright \dia\, 4-3 emission over most of B2. The greatest integrated \dia\, 4-3 intensity (not shown) is found towards E33, while additional strong \dia\, 4-3 emission is found towards E32 and the 850\,\micron\, continuum emission peak. A separate integrated intensity peak is found in northeast B2 (where we also find strong \ddia\, 3-2 emission). Overall, while the \dia\, 4-3 emission is generally confined to areas where there is also continuum emission, the strongest \dia\, 4-3 emission is found away from the brightest continuum contours. A comparison between Figure \ref{fig:n2h} and the \dia\, 1-0 integrated intensity in Figure \ref{fig:n2d}b (greyscale) reveals a significant discrepancy between the locations of strong \dia\, 1-0 and 4-3 emission in Oph B. We probe the possible causes of the offset, including the excitation and opacity of the lines, further below. 



\section{Analysis}
\label{sec:analysis_d}

\subsection{Line fitting}
\label{sec:line_fit}

The rotational transitions of the linear molecular ions \dia\, and \ddia\, contain hyperfine structure, which is dominated by the interactions between the molecular electric field gradient and the electric quadrupole moments of the two nitrogen nuclei \citep{dore04}. Rotational transitions at higher $J$ contain greater numbers of hyperfine components at smaller frequency intervals, i.e., the satellite components are found more closely spaced and can overlap significantly. In particular, the \dia\, 4-3 and the \ddia\, 3-2 transitions contain 21 and 25 hyperfine components, respectively, with relative line strengths $\gtrsim 0.001$ \citep{pagani09,gerin01}. 

For \dia\, 4-3, we calculate the relative line strengths $s_{J\,F_1\,F \rightarrow J^\prime\,F_1^\prime\,F^\prime}$ from the Einstein A coefficients $A_{J\,F_1\,F \rightarrow J^\prime\,F_1^\prime\,F^\prime}$ and frequencies $\nu_{J\,F_1\,F \rightarrow J^\prime\,F_1^\prime\,F^\prime}$ given by \citet{pagani09}, using

\begin{equation}
A_{J\,F_1\,F \rightarrow J^\prime\,F_1^\prime\,F^\prime} = \frac{64\pi^4}{3hc^2} \mu^2 
	\nu_{J\,F_1\,F \rightarrow J^\prime\,F_1^\prime\,F^\prime}^3  \frac{J}{[F]} 
	s_{J\,F_1\,F \rightarrow J^\prime\,F_1^\prime\,F^\prime}
\end{equation}

\noindent where $\mu = 3.4$\,D is the permanent electric dipole moment, $J$ is the upper rotational quantum number and $[F] = 2F + 1$ accounts for the degeneracy of the hyperfine states. For the \dia\, 4-3 transition, fourteen components at 8 distinct frequencies make up the main centroid emission, and lie within 0.4\,\kms\, of each other.  At low optical depths the line profile does not differ substantially from a Gaussian profile, while the relative intensities of the satellite lines to the main line component remain small even at large opacities ($\sim 20$\,\%\, at $\tau \sim 10$). Nevertheless, the satellite components of the line are visible towards several pixels in Oph B (identified as locations A through C on Figure \ref{fig:n2h}), largely near the embedded protostar E33. In some locations the line structure is confused with a significant blue emission shoulder, which is not due to any hyperfine structure, but which may be associated with an outflow stemming from either E33 or E32 \citep[][White {\it et al.} 2010, in preparation]{kamazaki03}. We were thus able to fit well the hyperfine structure towards only one location in the map, and constrain the excitation temperature, $T_{ex}$, and opacity, $\tau$, towards two others. 

We fit the hyperfine structure with a custom Gaussian hyperfine structure fitting routine written in {\sc IDL}, described in detail in \one, assuming equal excitation temperatures $T_{ex}$ for each hyperfine component. The line fitting routine does not take into account the significant overlap between hyperfine components in both the \dia\, and \ddia\, transitions, but we find the total \dia\, line opacities are still small where the satellite structure is seen ($\tau \sim 2-3$) and we therefore do not expect line overlap to introduce substantial errors. 




At low opacities, we cannot solve for $\tau$ and $T_{ex}$ independently. Given a reasonable estimate for $T_{ex}$, however, we may attempt to fit the line without visible satellite structure. In \two, we determined $T_{ex}$ for \dia\, 1-0 emission across Oph B, and to first order can expect the same $T_{ex}$ to describe the excitation of all rotational levels of \dia. This assumption was found to be accurate in a recent study of \dia\, and \ddia\, emission in L1544, where \citet{caselli02kin} were able to determine separately the excitation temperatures for \dia\, 1-0, 3-2 and \ddia\, 2-1 and 3-2, and found they were consistent within uncertainties. We find, however, that while most of the data can be fit reasonably well using the \dia\, 1-0 $T_{ex}$ values, visual inspection showed that a simple Gaussian fit the data equally well, or significantly better, in many cases, and the line opacities were not well-constrained. 

Because multiple components of similar line strength overlap to form the main \dia\, 4-3 component, a Gaussian fit will overestimate the line width, while the returned $v_{LSR}$ will be shifted relative to a full HFS fit. The shift in $v_{LSR}$ can be corrected ($\Delta v_{LSR} = 0.036$\,\kms), and at low opacities, the increase in $\Delta v$ is small ($\lesssim 10$\,\% for $\tau < 1$, up to $\sim 25$\,\% for $\tau \sim 2$). In the following sections, we thus discuss only the results of the Gaussian fitting of the \dia\, 4-3 emission unless otherwise stated. 

No hyperfine structure was seen above the rms noise in \ddia\, 3-2 emission across Oph B. We are therefore unable to fit well the \ddia\, 3-2 hyperfine structure solving for both $\tau$ and $T_{ex}$. Here, however, equating the \ddia\, $T_{ex}$ with that found for \dia\, 1-0 emission resulted in better constraints on the opacity. The hyperfine fit, using the relative component strengths and velocities from \citet{gerin01}, is thus more informative than a simple Gaussian fit, and we discuss the results from the full HFS fit in the following sections.


To properly calculate the line width, $\Delta v$, we subtracted in quadrature the resolution width, $\Delta v_{res}$ (0.1\,\kms\, for all species), from the observed line width, $\Delta v_{obs}$, such that $\Delta v = \sqrt{\Delta v_{obs}^2 - \Delta v_{res}^2}$. The uncertainties reported in the returned parameters are those determined by the fitting routine, and do not take the calibration uncertainty into account. The calibration uncertainty does not affect the uncertainties returned for $v_{\mbox{\tiny{LSR}}}$ or $\Delta v$. The excitation temperature, column densities and fractional abundances discussed further below, however, are dependent on the amplitude of the line emission, and are thus affected by the absolute calibration uncertainty.


The \hd\, $1_{11} - 1_{10}$ emission line has no hyperfine structure. We consequently fit the emission at each pixel with a single Gaussian (again in {\sc idl}) to determine the line amplitude, $v_{LSR}$ and $\Delta v$ across Oph B2. 

\subsection{Centroid velocity and line widths}
\label{sec:vel}

Table \ref{tab:tracers} lists the mean, rms, minimum and maximum values of $v_{LSR}$ and $\Delta v$ for \dia\, 4-3, \ddia\, 3-2, and \hd\, $1_{11} - 1_{10}$, along with the same results from \amm\, (1,1) and \dia\, 1-0 presented in \one\, and \two\, for comparison. 

We show in Figure \ref{fig:n2d_v} the \ddia\, \vlsr\, and $\Delta v$ over Oph B2. We only plot values where the S/N ratio of the peak line intensity was $\geq 5$. The mean \vlsr\, across B2 is 4.02\,\kms\, with an rms variation of 0.19\,\kms. Lower \vlsr\, values are found towards the western and northeastern regions of the Core, with the maximum \vlsr\, $= 4.44$\,\kms\, in the east towards the \amm\, (1,1) clump B2-A7 (identified in \one\, and labelled on Figure \ref{fig:n2d_v}; coincident with the \dia\, 1-0 clump B2-N6 from \two). The variation in \vlsr\, does not suggest global rotation of B2. The \ddia\, line velocities shift between 3.7\,\kms\, and 4.4\,\kms\, over $\sim 40$\arcsec, or $\sim 4800$\,AU, to the west of E32. Nearby, the line \vlsr\, shifts from 3.7\,\kms\, to 4.2\,\kms\, on either side of the western embedded protostar. 

The mean \ddia\, 3-2 line width $\Delta v = 0.53$\,\kms\, with an rms variation of 0.20\,\kms. In general, the widest line widths are found near E32 and the western protostar \citep[Oph-emb5,][]{enoch09}, to a maximum $\Delta v = 1.36$\,\kms\, in western B2 (but note the largest $\Delta v$ values are found only $\sim 45$\arcsec\, to the northwest of some of the narrowest $\Delta v$ values). The locations of large line widths correspond well with the locations of dramatic $v_{lsr}$ shifts, described above. Away from the protostars, only small variations in the relatively narrow line widths are seen, notably towards the northeast and southwest, with $\Delta v \sim 0.3$\,\kms. Two objects are associated with localized narrow $\Delta v$. An 850\,\micron\, continuum clump \citep[162725-24273, ][]{jorgensen08}, shown on the image, was identified where the southwestern narrow line widths are found. In the east, B2-A7 \amm\, is also associated with a $\Delta v$ minimum of 0.36\,\kms. This is similar to the clump line width in \dia\, 1-0 emission at 18\arcsec\, resolution, but nearly a factor of two larger than found in \dia\, 1-0 interferometer observations of the clump presented in \two. 

In Figure \ref{fig:both_all}a and Figure \ref{fig:both_all}c we show the \vlsr\, determined through Gaussian fitting of the \hd\, $1_{11} - 1_{10}$ and \dia\, 4-3 lines, respectively. Visual inspection of the \vlsr\, values shows that the \hd\, emission is similar kinematically to the \ddia\, 3-2 values in Figure \ref{fig:n2d_v}a, while a small shift in \vlsr\, of $\sim + 0.2$\,\kms\, is seen in \dia\, 4-3 emission towards Elias 32 and the 850\,\micron\, continuum emission peak and co-located 850\,\micron\, continuum emission peak (co-located with the continuum clump B2-MM8). The shift is small but significant given the otherwise excellent agreement in \vlsr\, between the different tracers over B2 (see Table \ref{tab:tracers}). 

We show in Figure \ref{fig:both_all}b and Figure \ref{fig:both_all}d the $\Delta v$ of the \hd\, and \dia\, lines. The \hd\, line widths are large, with a mean $\Delta v = 0.74$\,\kms, and do not vary greatly over Oph B2 ($\Delta v$ rms = 0.19\,\kms). Slightly greater line widths are found in the western half of B2. Gaussian fits to the \dia\, 4-3 data find the lines are slightly more narrow, with a mean $\Delta v = 0.69$\,\kms. As noted in \S\ref{sec:line_fit}, Gaussian fits likely overestimate the true \dia\, line width by $\sim 10 - 25$\,\%. Wider \dia\, lines are also found towards western Oph B2, while narrower lines are seen towards the submillimetre continuum emission peak and in B2's eastern half. \dia\, lines of average width are found towards the embedded protostars, but become wider at pixels directly offset from the infrared sources. In particular, directly to the east of E32 and E33, the \dia\, 4-3 line profiles show a significant blue shoulder with emission extending to \vlsr\, $\sim 2.5$\,\kms. Figure \ref{fig:n2h_spec}c shows the result of a two-component Gaussian fit to the line and blue shoulder at location C in Figure \ref{fig:n2h}. 

\subsection{\dia\, opacity and excitation temperature}
\label{sec:tex}

\begin{figure}
\plotone{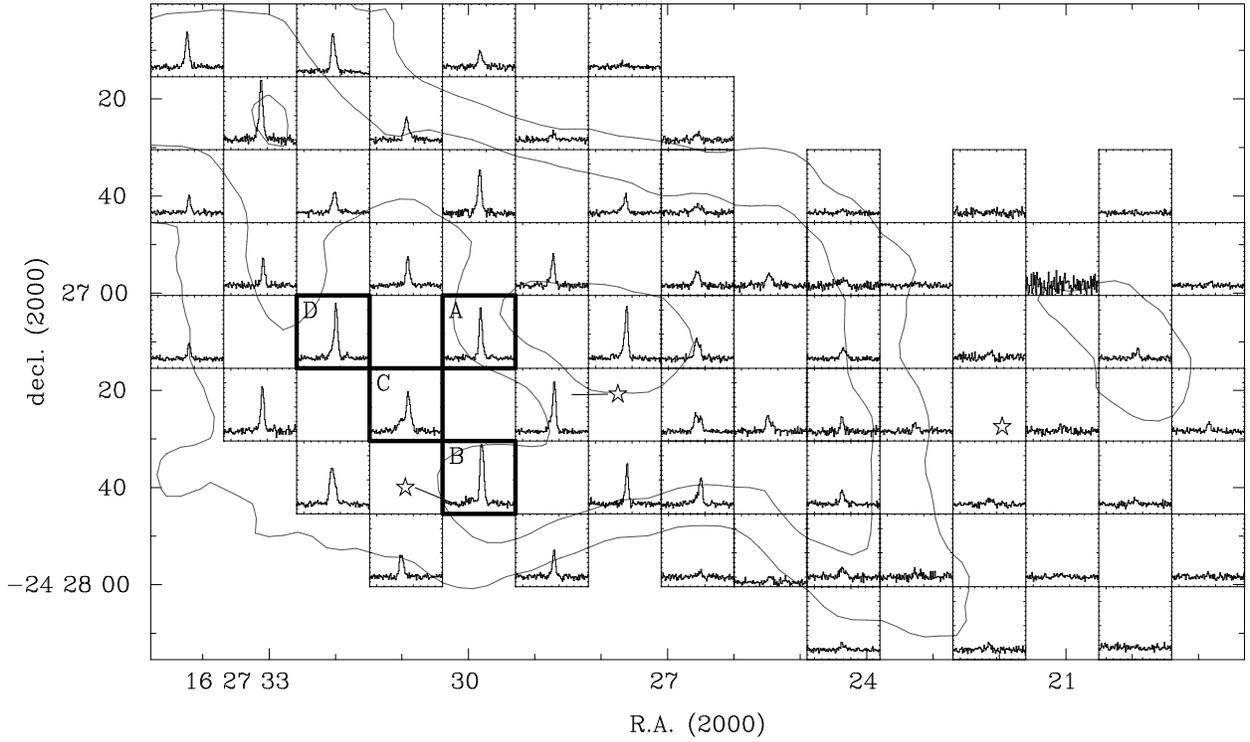} 
\caption{\dia\, 4-3 spectra observed with HARP at the JCMT towards Oph B2 with 15\arcsec\, FWHM. The grey contours, stars and velocity scale on the subplots are as in Figure \ref{fig:h2d}. The amplitude ranges from -0.3\,K to 2.2\,K ($T_{MB}$, assuming $\eta_{MB} = 0.6$). Four locations of interesting spectra, labeled A, B, C and D, are highlighted by dark outlines.}
\label{fig:n2h}
\end{figure}

Comparison of the \dia\, 1-0 and 4-3 line emission (see Figure \ref{fig:n2d}b and Figure \ref{fig:n2h}) shows that while in both cases emission is found over much of Oph B2, the integrated intensity distributions of the two transitions are quite different. It is thus unclear whether the excitation conditions for both transitions would be equal. 

\begin{figure}
\epsscale{0.6}
\plotone{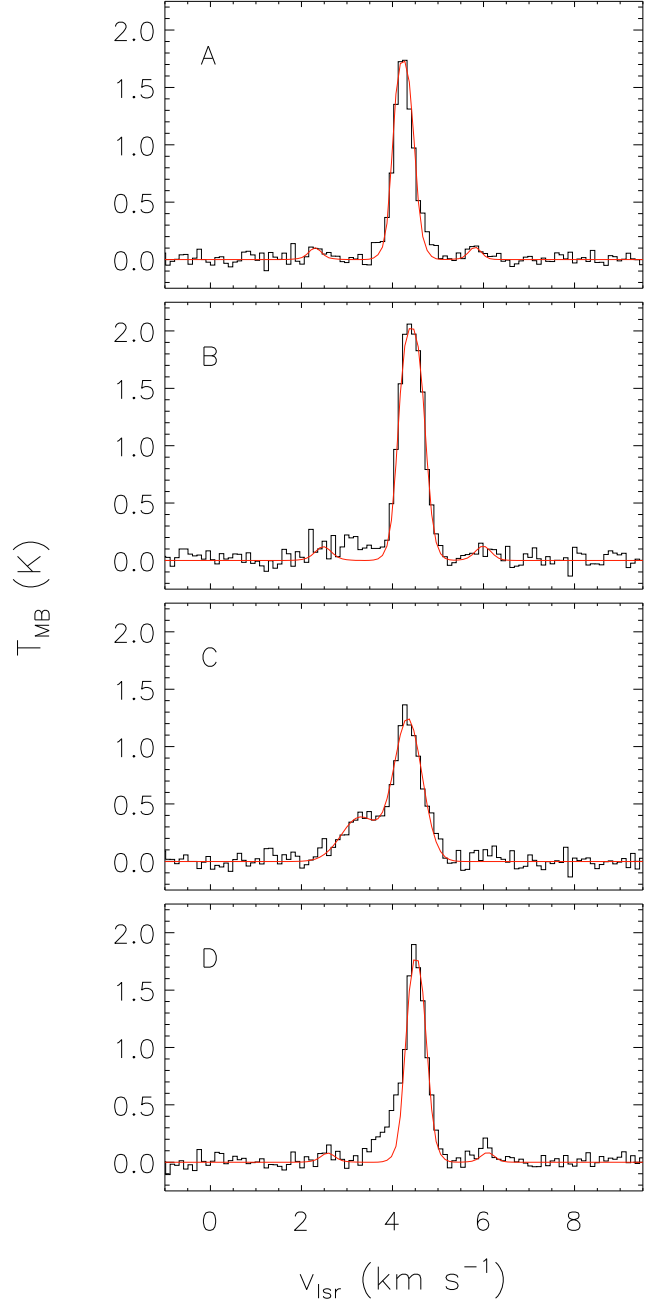} 
\caption{\dia\, 4-3 spectra towards four locations in Oph B2, labeled A, B, C and D in Figure \ref{fig:n2h}: 30\arcsec\, north of E33 (A), towards E33 (B), 21\arcsec\, north-east of E33 (C), and $\sim 42$\arcsec\, north-east of E33 (D). Overlaid on the spectra are fits to the full hyperfine structure of the \dia\, 4-3 line (A, B and D) and a two-component Gaussian fit (C).}
\label{fig:n2h_spec}
\epsscale{1}
\end{figure}

\begin{figure}
\plotone{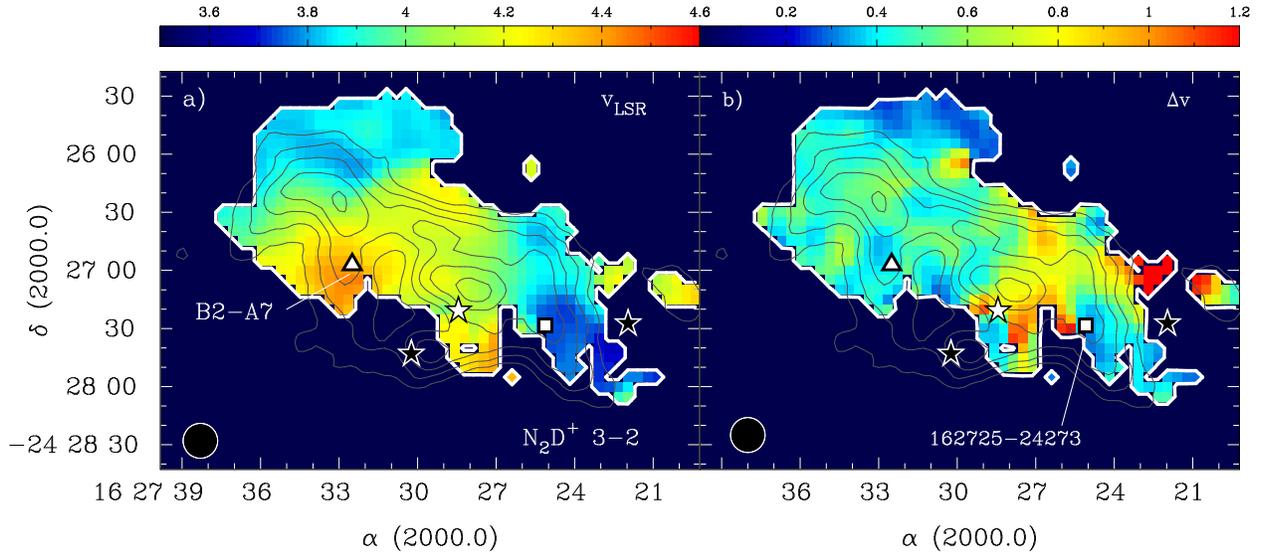} 
\caption[\ddia\, 3-2 fitted $v_{LSR}$ and $\Delta v$ in Oph B2]{a) \ddia\, 3-2 line velocity or $v_{LSR}$ in Oph B2. Colour scale is in \kms. In both plots, only results from pixels where the S/N of the peak line intensity is $\geq 5$ are shown, indicated by the solid white contour. Contours show 850\,\micron\, continuum emission at 15\arcsec\, (FWHM) resolution observed with the JCMT, in increments of 0.1\,Jy\,beam$^{-1}$. The 18\arcsec\, (FWHM) beam is shown at lower left. Stars indicate the positions of Class I protostars. The locations of the \amm\, clump B2-A7 and 850\,\micron\, continuum clump 16275-2422 are also shown. b) Fitted $\Delta v$ in Oph B2, contours as in a). Colour scale is in \kms. }
\label{fig:n2d_v}
\end{figure}

\begin{figure}
\plotone{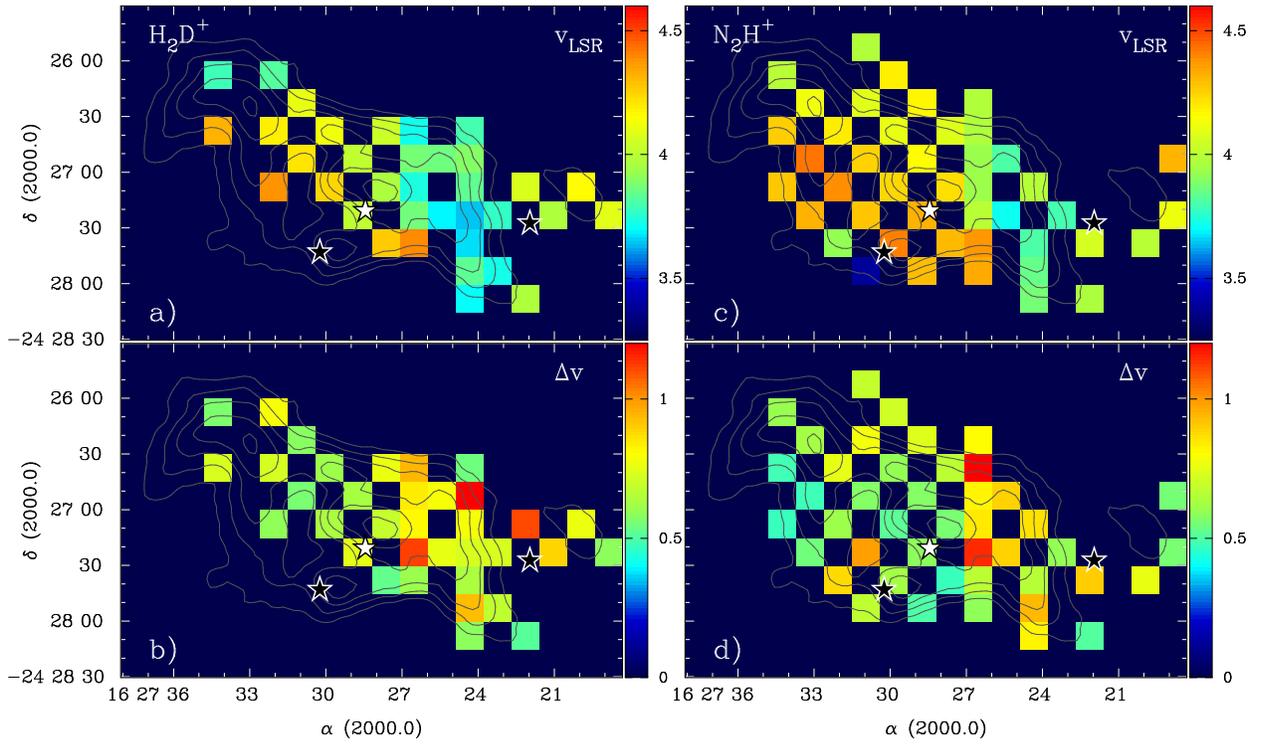} 
\caption[\hd\, $1_{11} - 1_{10}$ and \dia\, 4-3 \vlsr\, and $\Delta v$ in Oph B2]{a, c) \hd\, $1_{11} - 1_{10}$ and \dia\, 4-3 line velocity \vlsr\, in Oph B2, respectively. Colour scale is in \kms. We show only results from pixels where the S/N of the peak line intensity is $\geq 5$. Grey contours show 850\,\micron\, continuum emission as in Figure \ref{fig:n2d_v}. Stars indicate positions of embedded Class I protostars. b, d) \hd\, and \dia\, line width $\Delta v$ in Oph B2, respectively. Colour scale is in \kms.}
\label{fig:both_all}
\end{figure}

Towards several pixels, hyperfine structure is visible. These locations are labeled A through C on Figure \ref{fig:n2h}, and these spectra are again shown in Figure \ref{fig:n2h_spec}. Using the full hyperfine component fit (overlaid on the spectrum), we were able to determine reasonably well the excitation temperature and opacity of the \dia\, 4-3 line, $T_{ex, 4-3}$ and $\tau_{4-3}$ towards location A, with the resulting best fit total line opacity $\tau_{4-3} = 2.5 \pm 1$, with $T_{ex, 4-3} = 8 \pm 1$\,K. Both $T_{ex, 4-3}$ and $\tau_{4-3}$ are approximately consistent with values found in \two\, for \dia\, 1-0 emission at the same location, where $T_{ex, 1-0} = 7.3 \pm 0.1$\,K and $\tau_{1-0} = 3.1 \pm 0.6$. The brightest \dia\, 4-3 line emission, however, and additionally the maximum integrated intensity, is located towards E33 (location B in Figure \ref{fig:n2h}, directly south of A). The intensity ratio of the main \dia\, 4-3 component with an apparent red satellite (line strength only 2$\times$ the data rms noise level, but at the correct velocity and with a width that matches the main component, see Figure \ref{fig:n2h_spec}b) suggests that the \dia\, 4-3 opacity towards E33 remains moderately optically thick, $\tau_{4-3} \sim 3$, and the corresponding $T_{ex, 4-3} \sim 8$. The expected line profile given these values is overlaid on the spectrum in Figure \ref{fig:n2h_spec}. The integrated \dia\, 1-0 emission decreases significantly towards E33 (see Figure \ref{fig:n2d}b and \two). Additionally, the emission is confused by multiple velocity components which preclude robust fits to the 1-0 spectra, so we do not have a good estimate of the \dia\, 1-0 $T_{ex}$ at this location. At location D in Figure \ref{fig:n2h}, satellite emission is again seen, but here the hyperfine fits are confused by significant blue shoulder emission. The best-fit results give $T_{ex} \sim 7.5$\,K $- 10$\,K, with $\tau \sim 1 - 3$. We overlay the expected line profile for $T_{ex} = 8.5$\,K and $\tau = 2$ on spectrum D in Figure \ref{fig:n2h_spec}. Elsewhere, no hyperfine structure is visible in the \dia\, 4-3 line.

For opacity ratios between the \dia\, 4-3 and \dia\, 1-0 lines, $\tau_{4-3} / \tau_{1-0} \sim 0.3$ and similar line widths, the derived \dia\, column density, $N(\mbox{\dia})$, calculated using either transition following \citet{caselli02ion} (Equation A1), is equal if both $T_{ex, 1-0}$ and $T_{ex, 4-3} \sim 7$\,K. At lower opacity ratios, the required $T_{ex}$ also decreases. On average, the \dia\, 1-0 and 4-3 line widths agree, however we note that since we cannot fit well the hyperfine structure, the \dia\, 4-3 widths are an overestimate of the true value by $\sim 10-25$\,\%. This uncertainty propagates linearly into the column density calculation. With this caveat, an excitation temperature $T_{ex} \sim 7$\,K agrees well with the mean $T_{ex}$ determined from HFS fitting of the \dia\, 1-0 line in \two. We have no column density estimate towards E33 from \dia\, 1-0 emission. Near E32 and E33, however, the greater relative \dia\, 4-3 opacity requires either a higher common $T_{ex} \sim 10-11$\,K to obtain similar $N(\mbox{\dia})$ estimates, which does not agree with our \dia\, 1-0 $T_{ex}$ measurement, or the excitation of the \dia\, 4-3 line must be greater than that of the 1-0 line by $\sim 2-3$\,K. This moderate increase in excitation temperature of the higher excitation \dia\, line is not ruled out by our data. Detailed modeling of the \dia\, emission near the embedded protostars using radiative transfer codes would help disentangle the excitation and opacity effects. 


\subsection{Column density and fractional abundance}

We next discuss the column densities of \ddia\, and \hd\, derived from the observed line emission. The \dia\, column density was calculated in \two.  

\subsubsection{\ddia}
\label{sec:nn2d}

\begin{figure}
\plotone{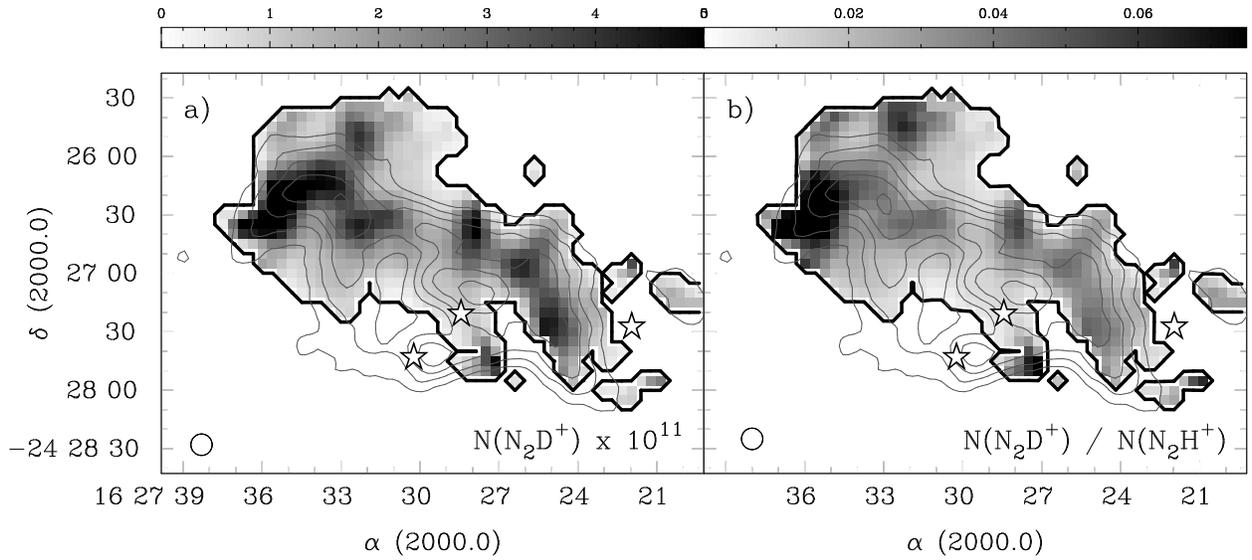} 
\caption[Total \ddia\, column density $N(\mbox{\ddia})$ and fractional abundance $X(\mbox{\ddia})$ in Oph B2]{a) Total \ddia\, column density $N(\mbox{\ddia})$ (cm$^{-2}$) in Oph B2 determined using Equation \ref{eqn:column_d} (greyscale). Values shown have been divided by $10^{11}$. In both plots, only results from pixels where the S/N of the peak line intensity is $\geq 5$ are shown, indicated by the thick black contour. Grey contours show 850\,\micron\, continuum emission as in Figure \ref{fig:n2d_v}. The 18\arcsec\, (FWHM) beam is shown at lower left. Stars indicate the positions of Class I protostars. b) Fractional abundance of \ddia\, relative to \dia, $R_D$, in Oph B2. The scale has been truncated to show the variation in central B2; the maximum $R_D = 0.16$ towards the eastern tip.}
\label{fig:n2d_c}
\end{figure}

We calculate the total \ddia\, column density, $N_{tot}(\mbox{\ddia})$, from the integrated intensity of the 3-2 transition following \citet{caselli02} for optically thin emission:

\begin{equation}
N_{tot} = \frac{8 \pi W}{\lambda^3 A} \frac{g_u}{g_l} \frac{1}{J_\nu(T_{ex}) - J_\nu(T_{bg})} 
		\frac{1}{1 - \exp(-h\nu/kT_{ex})} \frac{Q_{rot}(T_{ex})}{g_l \exp(-E_l/kT_{ex})}
\label{eqn:column_d}
\end{equation}

\noindent where $W$ is the line integrated intensity, $A$ is the Einstein spontaneous emission coefficient, $g_u$ and $g_l$ are the statistical weights of the upper and lower states, respectively, and the equivalent Rayleigh-Jeans excitation and background temperatures are given by $J_\nu(T_{ex})$ and $J_\nu(T_{bg})$. The rotational energy $E_J = J (J+1) h B$ for linear molecules, where the rotation quantum number $J=2$ for the lower energy state and $B = 38554.717$\,MHz is the rotational constant \citep{sastry81}. The partition function $Q_{rot} (T_{ex}) = \sum_{J=0}^\infty (2J+1) \exp(-E_J/kT_{ex})$, and we use the integrated value, $Q_{rot} (T_{ex}) = k T_{ex} \,/\, (hB) + 1/3$. We assume that the excitation temperature $T_{ex}$ is the same for all rotational levels, and use the fitted \dia\, 1-0 $T_{ex}$ from \two\, as an estimate of the excitation temperature of the \ddia\, line at each pixel (as discussed in \S\ref{sec:line_fit}). All other parameters remaining equal, \showrev{at the average $T_{ex}$ of the \dia\, 1-0 line,} a decrease in $T_{ex}$ by 2\,K would increase the resulting \ddia\, column density by a factor $\lesssim 2$, while an increase in $T_{ex}$ by 2\,K would decrease the resulting \ddia\, column density by $\lesssim 0.5$. We show $N(\mbox{\ddia})$ across Oph B2 in Figure \ref{fig:n2d_c}a. 


We then calculate the fractional \ddia\, abundance per pixel, 

\begin{equation}
X(\mbox{\ddia}) = N(\mbox{\ddia}) / N(\mbox{H$_2$})
\end{equation}

\noindent where the H$_2$ column density, $N(\mbox{H$_2$})$, is calculated from 850\,\micron\, continuum emission. The 15\arcsec\, data were convolved to a final 18\arcsec\, FWHM resolution to match the convolved \ddia\, data. The H$_2$ column density is then given by $N(\mbox{H$_2$}) = S_\nu / [ \Omega_m \mu m_{\mbox{{\tiny H}}} \kappa_\nu B_\nu (T_d)] $, where $S_\nu$ is the 850\,\micron\, flux density, $\Omega_m$ is the main-beam solid angle, $\mu = 2.33$ is the mean molecular weight, $m_{\mbox{{\tiny H}}}$ is the mass of hydrogen, $\kappa_\nu$ is the dust opacity per unit mass at 850\,\micron, where we take $\kappa_\nu = 0.018$\,cm$^2$\,g$^{-1}$, and $B_\nu (T_d)$ is the Planck function at the dust temperature, $T_d$. As in \one\, and \two, we expect the gas and dust to be well-coupled in Oph B2, and set \showrev{$T_d = T_K$, the gas kinetic temperature determined from \amm\, (1,1) and (2,2) inversion line ratios.} We refer the reader to the detailed discussion of the uncertainties in the $N(\mbox{H$_2$})$ values presented in \two, but note that the derived H$_2$ column densities have uncertainties of factors of a few. Additionally, due to the negative artifacts introduced into the data from the observational chopping technique (again discussed in detail in \two), we limit the analysis to pixels where $S_\nu \geq 0.1$\,Jy\,beam$^{-1}$, though the rms noise level of the continuum map is $\sim 0.03$\,Jy\,beam$^{-1}$. For a dust temperature $T_d = 15$\,K, this flux level corresponds to $N(\mbox{H}_2) \sim 6 \times 10^{21}$\,cm$^{-2}$. 

The column density distribution is not similar to the the integrated intensity map shown in Figure \ref{fig:n2d}a. Figure \ref{fig:n2d_c}a shows several small peaks of $N(\mbox{\ddia})$ in B2 which largely correspond with local minima in the excitation temperature ($T_{ex} \sim 5$\,K compared with the mean $T_{ex} \sim 7$\,K, where the uncertainty in $T_{ex}$ is $\sim 0.4-0.6$\,K). The $N(\mbox{\ddia})$ peaks remain within the continuum contours but avoid maxima of continuum emission. We find a mean $N(\mbox{\ddia}) = 1.8 \times 10^{11}$\,cm$^{-2}$ in Oph B2 with an rms variation of $1.3 \times 10^{11}$\,cm$^{-2}$. The maximum $N(\mbox{\ddia}) = 7.1 \times 10^{11}$\,cm$^{-2}$ is found towards the northeast. If we calculate $N(\mbox{\ddia})$ assuming a constant $T_{ex} = 7$\,K, the distribution follows the integrated intensity but the mean value and variation remain the same.

The fractional \ddia\, abundance (not shown) follows a similar distribution to the \ddia\, column density. We find a mean $X(\mbox{\ddia}) = 8.2 \times 10^{-12}$ with an rms variation of $5.1 \times 10^{-12}$. The maximum $X(\mbox{\ddia}) = 2.9 \times 10^{-11}$ is again found towards northeast Oph B2. Within the continuum emission contours, the lowest {\it measured} \ddia\, abundances ($X(\mbox{\ddia}) \sim 2 \times 10^{-12}$) are found towards the continuum emission peak and B2-MM8 clump (emission towards the Elias 33 protostar was not above our S/N limit for analysis). 

\subsubsection{\hd}
\label{sec:nh2d}

To calculate the \hd\, column density, $N(\mbox{\hd})$, we first estimate the line opacity $\tau$ from the \hd\, $1_{11} - 1_{10}$ line emission using

\begin{equation}
\tau = - \ln \biggr[ 1 - \frac{T_{MB}}{J(T_{ex}) - J(T_{bg})} \biggl]
\label{eqn:tau_d}
\end{equation}

\noindent where we must assume a priori an excitation temperature $T_{ex}$ for the transition. An upper limit is given by the gas kinetic temperature $T_K$. For the pixels where the \hd\, S/N ratio $> 5$, we found in \one\, a mean $T_K = 14.5$\,K with an rms variation of 1.1\,K (slightly less than the mean $T_K = 15$\,K found for the entire Oph B Core in \one), with a minimum $T_K \lesssim 13$\,K towards the highest $N(\mbox{H$_2$})$ values. We note, however, that the critical density at which \amm\, is excited ($n_{cr} \sim 10^{3-4}$\,\cc) is lower than that of the \hd\, transition by an order of magnitude, and the \amm\, temperatures, or the variation in $T_K$, may not accurately reflect the conditions where \hd\, is excited. 


In their recent \hd\, study of several starless and protostellar cores, \citet{caselli08} found that the \hd\, $T_{ex}$ was similar to or slightly less than the core $T_K$ values, which were estimated through a variety of methods (e.g., continuum observations, \amm\, inversion observations as for B2, and multiple transition observations of molecules such as HCO$^+$). Six of the seven protostellar cores in their sample had excitation temperatures which ranged from 9\,K to 14\,K and similar densities to those calculated for B2 in \two. Based on these values, and to better compare our results with \citeauthor{caselli08}, we use a constant excitation temperature $T_{ex} = 12$\,K in Equation \ref{eqn:tau_d}. Given $T_{ex} = 12$\,K, we find a mean $\tau = 0.13$, with a minimum $\tau = 0.05$ and maximum $\tau = 0.28$, which agree well with the \citeauthor{caselli08} results in protostellar cores. 

\begin{figure}
\plotone{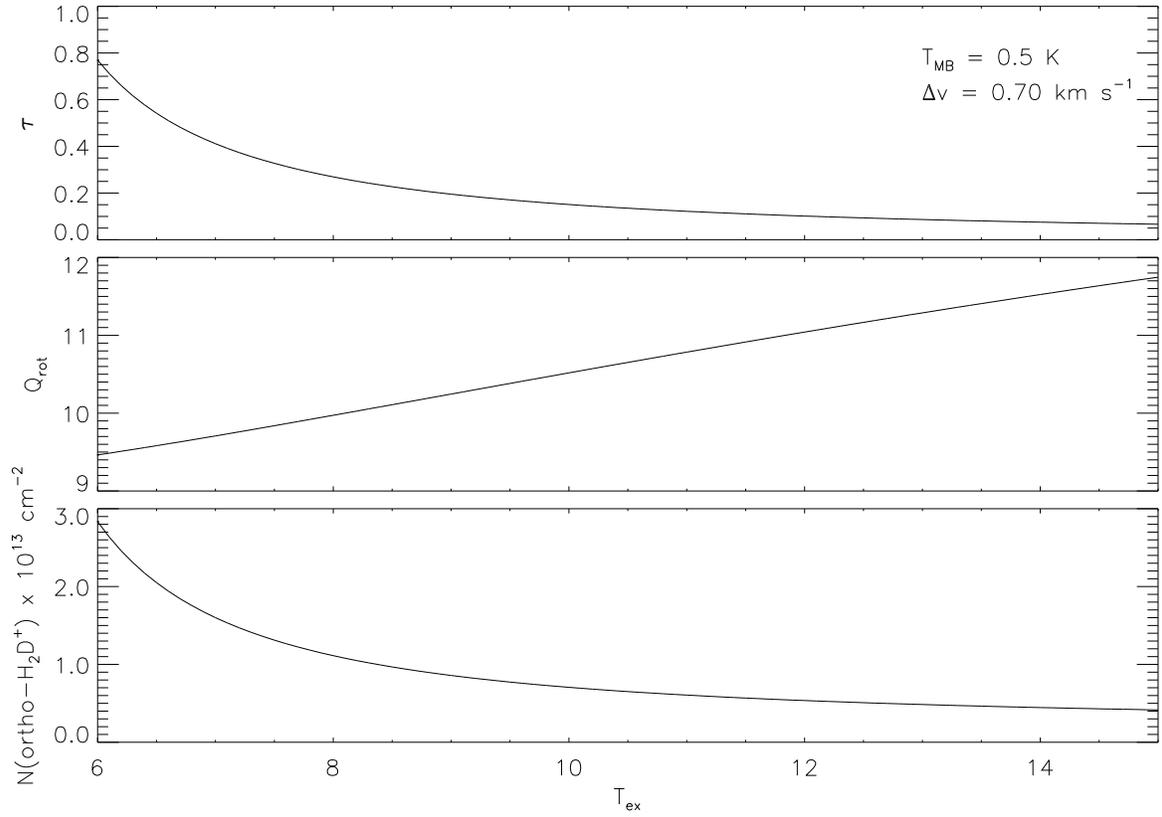} 
\caption[Impact of the assumed excitation temperature, $T_{ex}$, on returned \hd\, $1_{11} - 1_{10}$ line opacity (top), partition function $Q_{rot}$ (middle), and column density, $N(\mbox{ortho-\hd})$]{Impact of the assumed excitation temperature, $T_{ex}$, on returned \hd\, $1_{11} - 1_{10}$ line opacity (top), partition function $Q_{rot}$ (middle), and column density, $N(\mbox{ortho-\hd})$, given a line brightness temperature $T_{MB}$ and line width $\Delta v$ similar to observed mean values in Oph B2. At low $T_{ex}$, $\tau$ and $N(\mbox{ortho-\hd})$ vary strongly with $T_{ex}$, but the effect of changes in $T_{ex}$ at higher temperatures becomes small.}
\label{fig:tex_test}
\end{figure}

The total column density of ortho-\hd\, is then given by \citet{vastel06}

\begin{equation}
N(\mbox{ortho-\hd}) = \frac{8\pi}{\lambda^3A_{ul}}\frac{Q_{rot}(T_{ex})}{g_u} 
					\frac{\exp(E_u/kT_{ex})}{\exp(h\nu/kT_{ex})-1} \int \tau \mbox{d}v
\end{equation}

\noindent where $g_{u} = 9$, $E_u/k = 17.8$\,K, and $A_{ul} = 1.08 \times 10^{-4}$\,s$^{-1}$ for the $1_{11} - 1_{10}$ transition \citep{raman04}. The integral $\int \tau \mbox{d}v = \frac{1}{2}\sqrt{\pi/(\ln(2)} \tau \Delta v$. We calculate $Q_{rot}$ by reducing the \hd\, level structure to a 2-level system following \citeauthor{caselli08} Since the energy of the first excited state above ground is $E/k = 17.8$\,K and that of the second excited state is $E/k = 110$\,K, we expect this approximation to be valid in Oph B2 given the low temperatures determined  in \one. The partition function depends on $T_{ex}$, which we have estimated to be 12\,K. In Figure \ref{fig:tex_test}, we plot the variation in $\tau$, $Q_{rot}$ and $N(\mbox{ortho-\hd})$ with $T_{ex}$ given a line $T_{MB} = 0.5$\,K and $\Delta v = 0.7$\,\kms\, (typical values in B2), and have listed the returned values for $T_{ex}$ between 7\,K and 15\,K in Table \ref{tab:tex_test}. While the variation in returned parameters is fairly large with small $T_{ex}$ changes at low $T_{ex}$ values (i.e., factors $\sim 2$ or more around $T_{ex} \sim 7$\,K), at higher $T_{ex}$ the values vary less. For example, an increased $T_{ex} = 14$\,K rather than $T_{ex} = 12$\,K would result in a decreased $N(\mbox{ortho-\hd})$ by $\sim 20$\,\%, while a decreased $T_{ex} = 10$\,K would result in an increased $N(\mbox{ortho-\hd})$ by $\sim 30$\,\%. We thus assume an uncertainty of $\sim 25$\,\% in the $N(\mbox{\hd})$ values reported here. 

\showrev{To calculate the total \hd\, column density in B2, an estimate of the ortho- to para-\hd\, (o/p-\hd) ratio is needed. The ground state para-\hd\, transition occurs at a frequency of 1.37\,THz; this line is difficult to observe from the ground and may not be excited at the low temperatures found in star-forming cores. Without a direct measurement of the population of the para state, we look to chemical models to predict the o/p-\hd\, ratio, which show that the o/p-\hd\, ratio is directly dependent on o/p-H$_2$. \citet{walmsley04} find that the o/p-H$_2$ ratio increases to a steady-state value of $\sim 5 \times 10^{-4}$ at $T = 10$\,K and $n = 10^6$\,\cc, similar to the density and temperature determined for Oph B2 in \one\, and \two. The same models predict o/p-\hd\, $\sim 0.2-0.3$. Using Bonnor-Ebert sphere core models, \citet{sipila10} also predict o/p-\hd\, ratios $\lesssim 0.5$ at $T \sim 10-15$\,K in steady-state. Their models show that the o/p-\hd\, ratio can dramatically increase to values $\sim 2-3$ at $T \lesssim 10$\,K, but these temperatures are unlikely based on our analysis of Oph B2.  }

\showrev{It is not clear, however, whether steady-state models accurately represent conditions in star forming cores like B2. The dynamical timescale of a collapsing core can be much shorter than some of the chemical timescales, leading to large departures of free-fall model predictions from steady-state models \citep{flower06hd}. For example, \citeauthor{flower06hd} show that the steady-state model underpredicts the o/p ratio by up to an order of magnitude compared with a free-fall collapse model at gas densities $n \lesssim 10^6$\,\cc\, (the models are more consistent at higher densities). Consequently, it may not be unreasonable to expect variations in the o/p-\hd\, ratio across B2 as it evolves. Given this uncertainty in the o/p-\hd\, ratio, we limit discussion to $N(\mbox{ortho-\hd})$, but note that the total $N(\mbox{\hd})$ is likely greater than $N(\mbox{ortho-\hd})$ by a factor of a few.}



\begin{figure}
\plotone{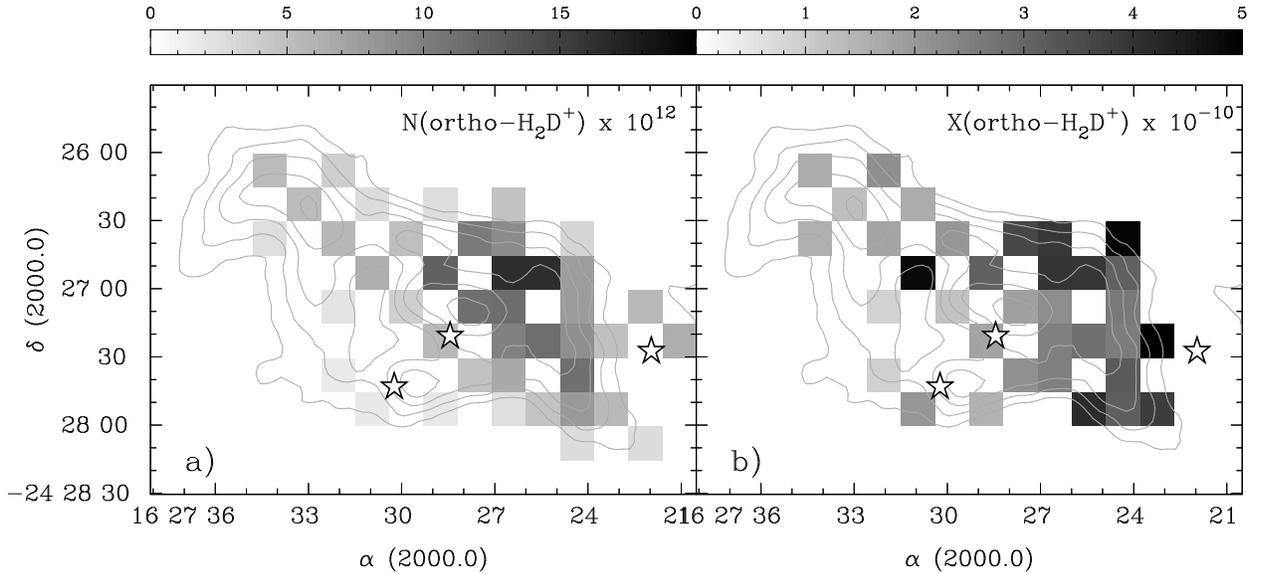} 
\caption[Ortho-\hd\, column density $N(\mbox{ortho-\hd})$ and fractional abundance $X(\mbox{ortho-\hd})$ in Oph B2]{a) Ortho-\hd\, column density in cm$^{-2}$ (greyscale). $N(\mbox{ortho-\hd})$ values have been divided by $10^{12}$. Grey contours show 850\,\micron\, continuum emission as in Figure \ref{fig:n2d_v}. b) Fractional ortho-\hd\, abundance $X(\mbox{ortho-\hd})$ (greyscale). Values have been divided by $10^{-10}$.}
\label{fig:h2d_c}
\end{figure}

We show in Figure \ref{fig:h2d_c} the resulting $N(\mbox{ortho-\hd})$ distribution across Oph B2. Including only pixels where the S/N $> 5$, we find the mean $N(\mbox{ortho-\hd}) = 1.4 \times 10^{13}$\,cm$^{-2}$ with an rms variation of $0.7 \times 10^{13}$\,cm$^{-2}$ over Oph B2. To this sensitivity limit, we find a minimum $N(\mbox{ortho-\hd}) = 4.1 \times 10^{12}$\,cm$^{-2}$. The maximum $N(\mbox{ortho-\hd}) = 3.3 \times 10^{13}$\,cm$^{-2}$ is found $\sim 30$\arcsec\, to the northwest of the 850\,\micron\, continuum peak. The largest ortho-\hd\, column densities are found in the western half of Oph B2, with moderate $N(\mbox{ortho-\hd})$ values extending to the northeast.  

We next calculate $X(\mbox{ortho-\hd})$ as described for \ddia\, above, and show in Figure \ref{fig:h2d_c}b the $X(\mbox{ortho-\hd})$ distribution across B2. Similarly to the column density distribution, the largest \hd\, abundances, $X(\mbox{ortho-\hd}) \sim 4-5 \times 10^{-10}$, are found in the west. A local $X(\mbox{ortho-\hd})$ minimum of $\sim 2 \times 10^{-10}$ is found towards the central continuum emission peak, with similar abundances towards the northeast. 

\subsection{Using \hd, \dia\, and \ddia\, to determine $T_K$}
\label{sec:tk}

The molecular weight of \hd\, ($m_{mol} = 4.01$\,m$_H$) is much less than that of \dia\,  or \ddia\, (29.01\,m$_H$ and 30.02\,m$_H$, respectively). If we assume that the \hd\, $1_{11} - 1_{10}$, \dia\, 4-3 or \ddia\, 3-2 transitions are excited in the same environment within Oph B2, we would expect each to trace the same core motions, and be described by the same kinetic temperature $T_K$.  This co-location would imply that the non-thermal line widths for the two species should be equal, where $\sigma_{NT}^2 = \sigma_{obs}^2 - \sigma_{th}^2$ . Given equal $T_K$, the \hd\, thermal line width, $\sigma_{th} = (kT_k/m_{mol})^{1/2}$, will be measurably larger than that of \dia\,  due to its smaller molecular weight. At $T_K = 12$\,K, for example, $\sigma_{th} = 0.06$\,\kms\, for \ddia\, and 0.16\,\kms\, for \hd. We can use this difference in thermal line widths to determine the kinetic gas temperature by equating the non-thermal line widths and solving for $T_K$ (e.g., in the case of \dia): 

\begin{equation}
T_K = \frac{1}{k_B}\biggl(\frac{1}{m_{\mbox{\tiny{\hd}}}} - \frac{1}{m_{\mbox{\tiny{\dia}}}}\biggr)^{-1}
			[\sigma_{obs}^2(\mbox{\hd}) - \sigma_{obs}^2(\mbox{\dia})]
\label{eqn:tk_d}
\end{equation}

Mean temperatures of $9.2 \pm 0.2$\,K, in good agreement with temperatures derived from \amm\, alone, were derived in this manner towards a number of dense cores by \citet{fuller93} using the molecular species HC$_3$N and \amm. The largest source of error in the derived $T_K$ likely arises from the assumptions listed above rather than from the propagation of the small uncertainties on the observed line widths. 

Overall, we expect the comparison of \hd\, and \ddia\, to be most accurate, given the similar critical densities of the transitions, the fact that \ddia\, must form from \hd, and the similar distribution of integrated intensity (in contrast with the poor correlation between \hd\, and \dia\, 4-3 emission; see Figures \ref{fig:n2d}a and \ref{fig:h2d}). The different distributions of \hd\, and \dia\, 4-3 in Oph B2 suggest that these two species are not excited in the same environment in the core. Additionally, since our estimates of the \dia\, line widths are not accurate (given the unresolved hyperfine structure), we will focus only on \ddia\, for this analysis.

\begin{figure}
\plotone{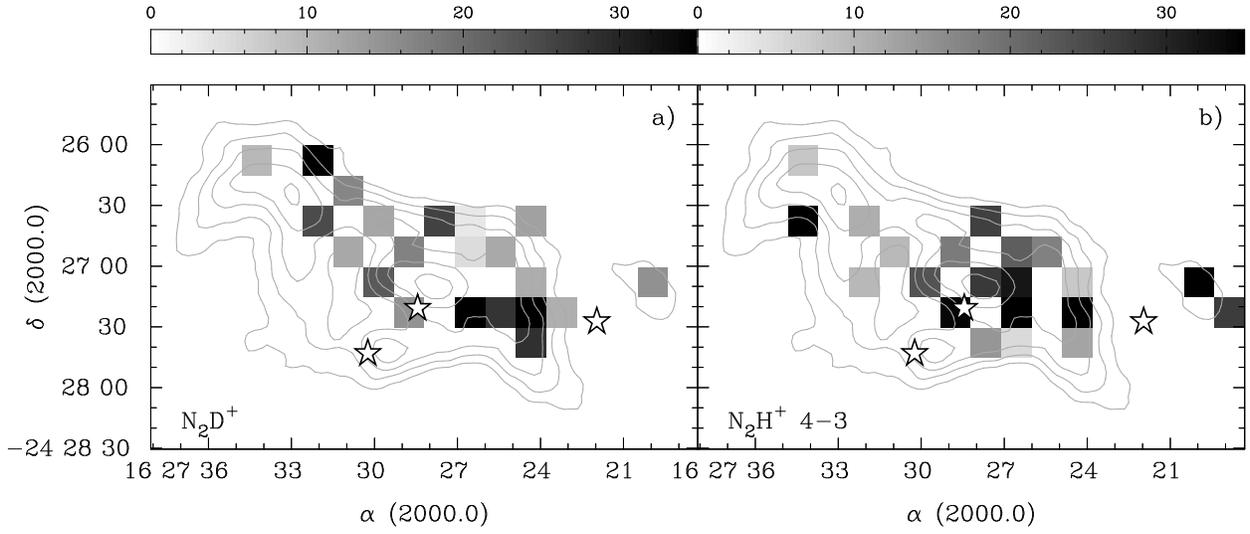} 
\caption[$T_K$ determined using Equation \ref{eqn:tk_d} for \hd\, and \ddia\, emission and \hd\, and \dia\, 4-3 emission]{a) $T_K$ determined using Equation \ref{eqn:tk_d} for \hd\, and \ddia\, emission. Note that a negative $T_K$ was found for the blank pixel towards the submillimetre continuum emission peak because the observed \hd\, $\Delta v > $ \ddia\, $\Delta v$.  Grey contours show 850\,\micron\, continuum emission as in Figure \ref{fig:n2d_v}. b) $T_K$ determined using Equation \ref{eqn:tk_d} for \hd\, and \dia\, 4-3 emission. }
\label{fig:tk}
\end{figure}

We show in Figure \ref{fig:tk} the temperatures calculated based on Equation \ref{eqn:tk_d} for \hd\, and \ddia\, 3-2. Only pixels where the signal-to-noise ratio of both lines was $> 5$ are shown. The derived $T_K$ values span a large range, from unphysical negative values ($\sim -11$\,K) to unrealistically large values ($\sim 75$\,K). Using the mean $\Delta v$ values of \hd\, and \ddia\, from Table \ref{tab:tracers}, we calculate a mean $T_K = 16$\,K, similar to that determined through \amm\, analysis. Multiple pixels with $T_K \sim 12$\,K are found surrounding the central continuum emission peak. Negative temperatures, however, are found associated with the continuum  peak and associated clump B2-MM8.

\begin{figure}
\plotone{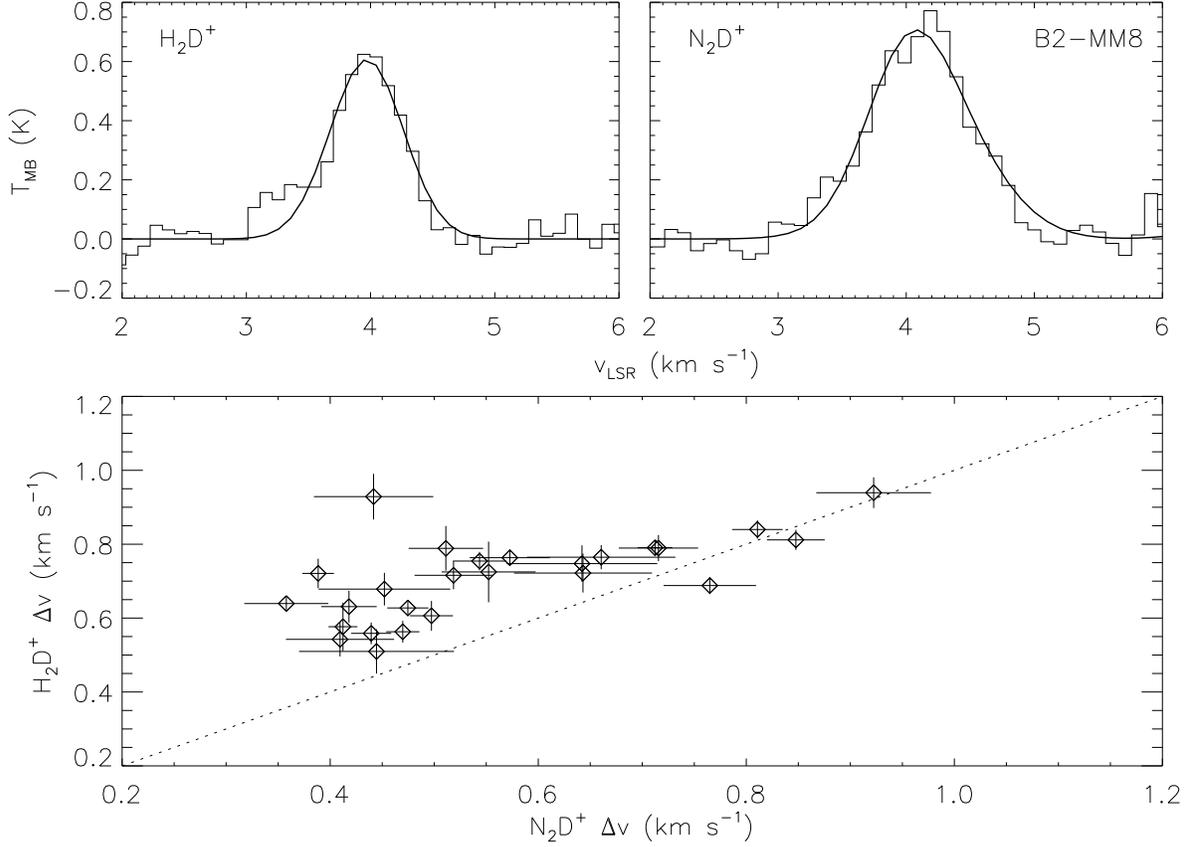} 
\caption[Comparison of returned \hd\, $1_{11} - 1_{10}$ and \ddia\, 3-2 $\Delta v$ in Oph B2]{Top left: \hd\, $1_{11} - 1_{10}$ spectrum towards B2-MM8 overlaid with a Gaussian fit. Top right: \ddia\, 3-2 spectrum towards B2-MM8 overlaid with a hyperfine structure fit (note no individual hyperfine components are visible due to their small separation in velocity and the relatively large line width). Bottom: Returned \hd\, line widths $\Delta v$ versus \ddia\, $\Delta v$. Each point represents a 15\arcsec\, pixel. An equality relation is shown by the dotted line. Despite appearing wider, in most locations the \ddia\, $\Delta v$ is narrower than that of \hd\, due to the presence of overlapping hyperfine lines. The two points where \ddia\, $\Delta v > $ \hd\, $\Delta v$ are located at and immediately adjacent to the 850\,\micron\, continuum peak and B2-MM8. At these locations, temperatures calculated via Equation \ref{eqn:tk_d} become negative and unphysical. Uncertainties shown are returned from the line fitting routines. Uncertainties in the line widths are shown.}
\label{fig:n2d_h2d_mm8}
\end{figure}

From Equation \ref{eqn:tk_d}, we calculate negative $T_K$ values when the observed \hd\, line widths are less than those of \ddia\, 3-2. We show in Figure \ref{fig:n2d_h2d_mm8} (bottom) $\Delta v$ derived for \hd\, and \ddia\, emission (data values in 15\arcsec\, pixels plotted), along with the spectra and relative fits towards B2-MM8 (top). Over most of B2, \ddia\, line widths are narrower than \hd, and thus $T_K > 0$\,K. The negative temperatures derive from two adjacent pixels towards B2-MM8 where the \ddia\, $\Delta v > \Delta v$ of \hd. Figure \ref{fig:n2d_h2d_mm8} shows non-Gaussianity in the spectra of both species towards B2-MM8, which complicates the calculation of a single line width to describe the emission.


Why would we find greater \ddia\, $\Delta v$ than \hd\, $\Delta v$ only towards the central continuum peak? One possibility is suggested by the spectra in Figure \ref{fig:n2d_h2d_mm8}, where additional emission in the \hd\, spectrum is clearly seen blue-shifted relative to the main Gaussian peak. For the \hd\, line, the best-fit Gaussian does not include the blue line shoulder. Slight blue asymmetry is also visible in the \ddia\, line, but the distinction between the main component and the shoulder is less obvious. It is possible that the shoulder emission of \ddia\, 3-2 is spread over a larger range in \vlsr\, due to the contribution of multiple hyperfine components to the observed emission (which would each contain emission at the central line velocity and at the shoulder velocity), broadening the line overall. The best fit to the \ddia\, line thus cannot distinguish well between the main and shoulder emission, resulting in an artificially broad $\Delta v$. 


\section{Discussion}
\label{sec:discussion}

\subsection{Line widths and density}
\label{sec:linewidths}

The mean line widths listed in Table \ref{tab:tracers} show that on average, the observed $\Delta v$ for a given molecular line decreases with increasing critical line density, suggesting that the gas in B2 becomes more quiescent at higher densities (omitting \dia\, 4-3, where $\Delta v$ is an overestimate). For an average gas temperature $T_K = 15$\,K, the resulting non-thermal line contribution decreases from $\sigma_{\mbox{\tiny NT}} = 0.36$\,\kms\, for \amm\, (1,1) emission to $\sigma_{\mbox{\tiny NT}} = 0.22$\,\kms\, for \ddia\, 3-2 emission (for $n_{cr} \sim 10^{3-4}$\,\cc\, to $n_{cr} \sim 8 \times 10^5$\,\cc, respectively). Comparing with the thermal sound speed at 15\,K, where $c_s = \sqrt{kT_K/(\mu m_{\mbox{\tiny H}})}$, we find an average $\sigma_{\mbox{\tiny NT}}\,/\,c_s = 1.6$ for \amm\, and $\sigma_{\mbox{\tiny NT}}\,/\,c_s = 1.1$ for \ddia\, 3-2 emission. This comparison shows that despite the decrease in non-thermal motions, the non-thermal motions are equal in magnitude with the sound speed, even at high density. A lower gas temperature would result in even larger magnitudes of $\sigma_{\mbox{\tiny NT}}$, while significantly higher temperatures are unlikely. 

This decrease in non-thermal motions at higher densities has been observed in starless cores \citep[e.g., ][]{pon09,tafalla04,lada03b68}.  Frequently in starless cores, however, the narrowest observed line widths are consistent with being due to nearly pure thermal motions, in contrast to the results found here. As $\sigma_{\mbox{\tiny NT}}$ values decrease, the ability for turbulent pressure to support the core against collapse is reduced, which impacts the stability of the core against its self-gravity. The narrow $\sigma_{\mbox{\tiny NT}}$ widths found in starless cores are often explained as the result of the dissipation of turbulent motions in the core centre \citep{goodman98}. If the non-thermal line widths in B2 are due to turbulent motions, the corresponding turbulent pressure ($P_{NT} = mn\sigma_{\mbox{\tiny NT}}^2$) is equal to or greater than the thermal pressure ($P_T = nkT$). 

In \one, we discussed the source of wide \amm\, line widths in Oph B, and concluded that if the relatively large $\sigma_{\mbox{\tiny NT}}$ values are caused by turbulent motions, they could not be primordial (i.e., inherited from the parent cloud). This is because the damping timescale of such motions (with no external driving force) is approximately the dynamical time \citep{maclow04}, and the existence of protostars associated with B2 implies that B2 has likely existed, with high density gas, for at least a dynamical time. 

The wide lines in Oph B2 may, however, be due to more organized motions such as infall or outflow. None of the observed molecular lines are optically thick, and we therefore do not see the self-absorbed, asymmetric line profiles frequently used to infer the infall of gas within a core. Infall signatures were observed in CS, H$_2$CO or HCO$^+$ emission towards several locations in eastern B2 by \citet{andre07}, but not in central or western B2. \citet{gurney08} observed complicated CO (and CO isotopologues) line structures towards the B2 continuum peak, but were unable to distinguish infall conclusively. The analysis of line profiles is complicated by a substantial outflow discovered in CO emission and emanating from one of the two embedded YSOs in B2 \citep{kamazaki03} which extends over 10\arcmin\, ($\sim 0.4$\,pc) in size\footnote{JCMT Spring 2009 Newsletter, http://jach.hawaii.edu/JCMT/publications/newsletter/}. Upcoming results of CO, $^{13}$CO and C$^{18}$O 3-2 observations in Ophiuchus (White et al. 2010, in preparation) from the JCMT Gould Belt Legacy Survey \citep[GBLS, ][]{gbls} will be better able to study the infall and outflow motions in Oph B. In conjunction with these data, the process of how the outflow motions affect the dense gas may be studied. There is clear evidence, however, for protostellar influence on the \ddia\, line widths shown in Figure \ref{fig:n2d_v}b, as discussed in \S\ref{sec:vel}. Additionally, the blue shoulder emission seen in \dia\, and \ddia\, east of E32 and E33 coincides with the eastern CO lobe of the outflow, and is further evidence for the significant impact of the outflow on the dense gas. 

We also noted in \S\ref{sec:vel} that there is an offset in \vlsr\, of $\sim 0.2$\,\kms\, between \dia\, 4-3 and \hd\, emission towards B2-MM8 and the nearby embedded protostar Elias 32, with the \dia\, emission redshifted relative to the \hd\, emission. Over the rest of Oph B2, the correlation between the \vlsr\, of the two lines is good, suggesting the offset is real and localized. Comparison with \ddia\, emission at the same location shows that the \vlsr\, of the \hd\, and \ddia\, agree well, and it is consequently the \dia\, emission which is offset in velocity from the rest of the core gas. \dia\, 4-3 has a critical density approximately an order of magnitude greater than the \hd\, and \ddia\, lines discussed here. The offset in \vlsr\, may thus be probing a shift in the gas motions at extremely high densities in the core interior. 

\subsection{Trends in the deuterium fractionation}
\label{sec:chem}

\subsubsection{The \ddia/\dia\, ratio}

The deuterium fractionation, $R_D$, can be defined as the ratio between the column densities of a deuterated molecule and its hydrogen-bearing counterpart. Here, we define $R_D = N(\mbox{\ddia})/N(\mbox{\dia})$, and calculate $R_D$ in Oph B2 from $N(\mbox{\ddia})$ calculated in \S\ref{sec:nn2d} and $N(\mbox{\dia})$ calculated in \two. Figure \ref{fig:n2d_c}b shows $R_D$ across B2. The distribution is similar to the \ddia\, column density in Figure \ref{fig:n2d_c}a. The largest values of $R_D$, to a maximum $R_D = 0.16$, are found towards northeast B2, while towards the smaller scale maxima, we find $R_D \sim 0.1$ or less. We estimate we are sensitive to $R_D \sim 0.01 - 0.015$ based on the rms noise in the \ddia\, integrated intensity map ($3\sigma\, \sim 0.2$\,K\,\kms\, in the off-B2 pixels), and the mean \dia\, column density over Oph B2, $\langle N(\mbox{\dia}) \rangle = 5.9 \times 10^{12}$\,cm$^{-2}$, calculated in \two. We find a moderate mean $R_D = 0.03$ with an rms variation of 0.01 towards B2, comparable to results in Ori B9 towards two \dia\, emission peaks and a protostellar source \citep[$R_D = 0.03-0.04$, ][]{miett09}. While \citet{pagani09} find substantially greater central $R_D$ values through modelling of the starless core L183 ($R_D \sim 0.7$ at the core centre) which decrease to $\sim 0.06$ at $\sim 5000$\,AU, it is not clear how to compare their model of $R_D$ as a function of core radius to our observed column density ratio. \citet{fontani06} found an average $R_D \sim 0.01$ towards the high mass star forming region IRAS 05345+3157, which at high resolution resolved into two \ddia\, condensations each with $R_D = 0.11$ \citep{fontani08}. In a study of starless cores, \citet{crapsi05} find a range in $R_D$ between a lower limit of 0.02 to a maximum of 0.44 (which was found towards the Oph D Core). \citeauthor{crapsi05} also note that $R_D$ values  $ > 0.1$ are generally only found for cores with $N(\mbox{\dia}) > 10^{13}$\,cm$^{-2}$, which is approximately the maximum $N(\mbox{\dia})$ in Oph B2. A similar range in $R_D$ was found towards protostellar, Class 0 sources \citep{emprechtinger09}. On average, then, our results agree with previous results. Given our map of Oph B2, however, we can also probe the variation of $R_D$ across the core.

\begin{figure}
\plotone{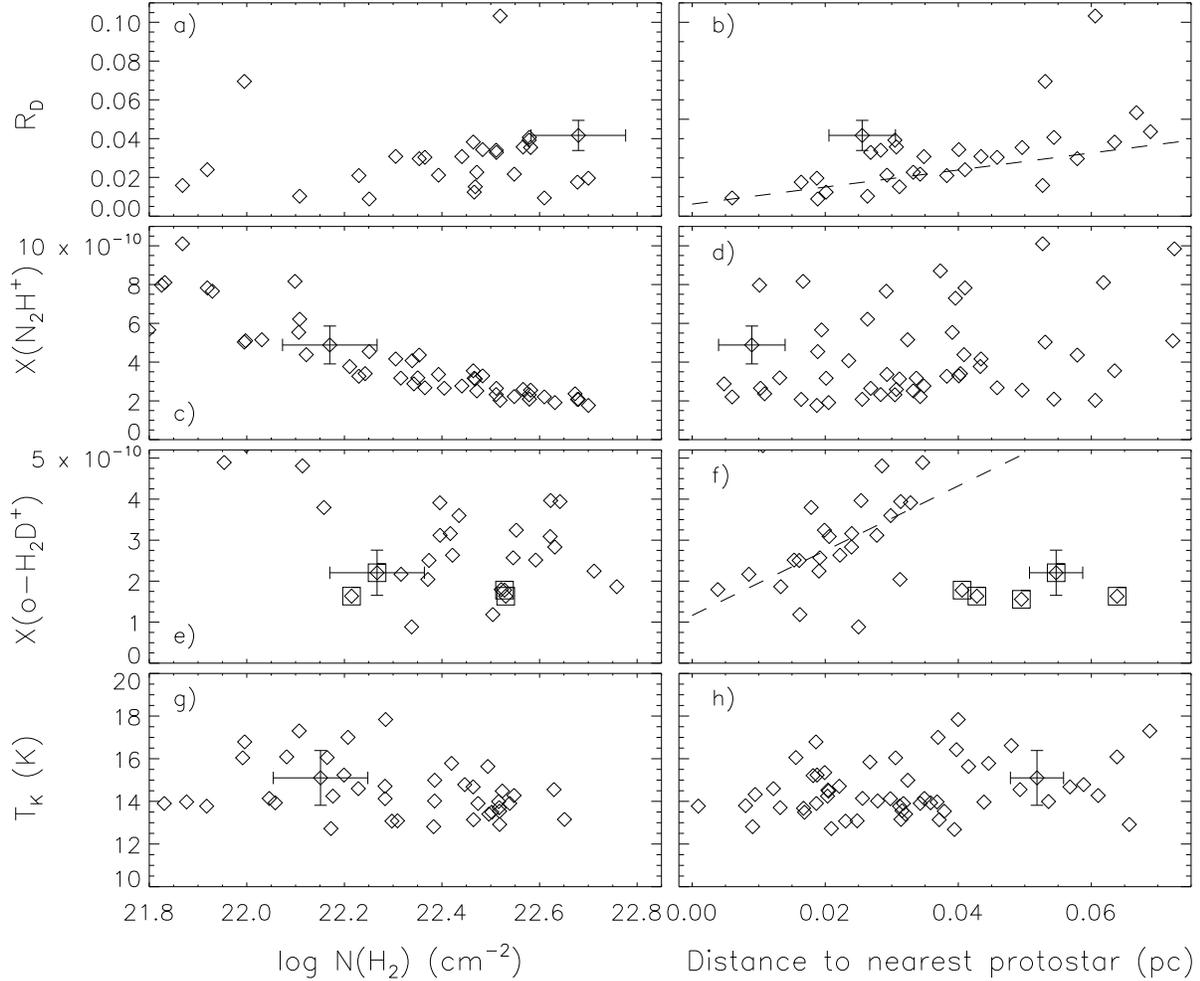} 
\caption{Variation in $R_D$, $X(\mbox{\dia})$, $X(\mbox{ortho-\hd})$ and $T_K$  with H$_2$ column density and the projected distance to the nearest embedded protostar in Oph B2. In each plot, points shown represent pixels matched to the FWHM beam of the respective datasets (15\arcsec\, for $T_K$ and \hd, and 18\arcsec\, for \dia\, and \ddia). Typical uncertainties are shown for a single pixel in each plot. An uncertainty of $\sim 20-30$\,\% in the $X(\mbox{ortho-\hd})$ values is not shown. Only pixels with S/N $> 5$ are plotted. In (a), no trend is found in $R_D$ versus $\log N(\mbox{H$_2$})$. The $R_D$ data are, however, consistent with an increasing deuterium fractionation with distance from an embedded source (b), albeit with significant scatter. As discussed in \two, a strong correlation is seen in $X(\mbox{\dia})$ versus $\log N(\mbox{H$_2$})$ (c), but no trend is seen with respect to projected protostellar distance (d). $X(\mbox{ortho-\hd})$ shows no trend with $\log N(\mbox{H$_2$})$ (e).  With respect to projected protostellar distance, the $X(\mbox{ortho-\hd})$ plot shows the data do not scatter about a central point (f), but instead show a trend of increasing $X(\mbox{\hd})$ to $d \sim 0.04$\,pc, and a smaller, constant \hd\, abundance at further distances. In both (e) and (f), the points identified with squares represent pixels in the northeast portion of the B core. A linear fit to the data, omitting the highlighted pixels, is shown by the dotted line. A linear fit to the data is shown by the dashed line. No clear correlation is found in $T_K$ versus $\log N(\mbox{H$_2$})$ (g) or projected protostellar distance (h). }
\label{fig:protall}
\end{figure}

In Figure \ref{fig:protall}, we show $R_D$ as a function of increasing $N(\mbox{H$_2$})$ (Figure \ref{fig:protall}a), and also as a function of increasing distance (in projection) from the nearest embedded protostar (Figure \ref{fig:protall}b). Each data point represents an 18\arcsec\, pixel (to match the beam FWHM) and pixels with S/N $< 5$ in either \dia\, or \ddia\, emission are omitted.  No clear trend is present in the deuterium fractionation with $N(\mbox{H$_2$})$. We find, however, that $R_D$ increases at greater projected distances from the embedded sources, as might be expected from the \ddia\, integrated intensity distribution. We fit a linear trend to the data, where $R_D = (-0.007 \pm 0.002) + (0.37 \pm 0.06) d\,\mbox{[pc]}$, but note that there is large scatter in the relation. An analysis of the individual \dia\, and \ddia\, abundances shows that the increasing $R_D$ trend with projected protostellar distance is dominated by a similar increase in $X(\mbox{\ddia})$ with projected protostellar distance, while $X(\mbox{\dia})$ is approximately constant over the range plotted, shown in Figure \ref{fig:protall}d. 

\subsubsection{\hd}

We next look at the distribution of the ortho-\hd\, abundance over Oph B2. In Figure \ref{fig:protall}, we show $X(\mbox{ortho-\hd})$ as a function of $N(\mbox{H$_2$})$ (Figure \ref{fig:protall}e), and also as a function of increasing projected protostellar distance (Figure \ref{fig:protall}f). We show data in 15\arcsec\, pixels, and omit pixels with S/N $< 5$. We find no trend in the fractional ortho-\hd\, abundance with H$_2$ column density (an increasing trend is found, but not shown, in the ortho-\hd\, column density, $N(\mbox{ortho-\hd})$, with $N(\mbox{H$_2$})$). The $X(\mbox{ortho-\hd})$ vs. $N(\mbox{H$_2$})$ relationship is not entirely a scatter plot, as the ortho-\hd\, abundance tends to increase with projected protostellar distance to a maximum $X(\mbox{ortho-\hd}) = 4.9 \times 10^{-10}$ at $d = 0.035$\,pc, with $X(\mbox{ortho-\hd}) \sim 1.8 \times 10^{-10}$ at larger distances. All data points at $d > 0.035$\,pc correspond to the emission found in northeast B2, while the rest show emission closer to the central continuum emission peak. If we fit a linear relationship to the data, omitting the northeast B2 points, we find $X(\mbox{ortho-\hd}) \times 10^{-10} = (1.2 \pm 0.5) + (79 \pm 23)  d\,\mbox{[pc]}$. 

There is also no clear trend in $N(\mbox{ortho-\hd})$ with $R_D$ where both \hd\, and \ddia\, are strongly detected across Oph B2, although both species are found with relatively large column densities north and west of the continuum peak. Given the small-scale structure found in the $N(\mbox{\ddia})$ distribution in B2, the spatial resolution and sampling of our \hd\, observations may not be high enough to discern a correlation. We note, however, that \citet{caselli08} also found no clear correlation between $R_D$ and $N(\mbox{ortho-\hd})$ in their sample of starless and protostellar cores (with the caveat that some $R_D$ values were derived using column density ratios of deuterated \amm\, and deuterated H$_2$CO with their undeuterated counterparts rather than $N(\mbox{\ddia})/N(\mbox{\dia})$). In particular, high values of $R_D$ and relatively low $N(\mbox{ortho-\hd})$ were found towards two objects in Ophiuchus (Oph D and 16293E). 

\subsection{What is affecting the deuterium fractionation in Oph B2?}

The largest deuterium fractionation values in Oph B2 are not found towards the continuum emission peak, but instead avoid the continuum peak and nearby protostars. The results of the previous section indicate that the distance to the nearest protostar is the dominant parameter impacting the deuterium fractionation in Oph B2. \showrev{We next discuss several mechanisms by which the embedded sources are most likely to impact the deuterium fractionation distribution in B2, including temperature variation in the gas, liberation of CO from dust grains, the impact of protostellar outflows, and possible increase in the ionization fraction due to x-rays from the protostars. }

\subsubsection{Temperature and CO depletion}

\showrev{Temperature can have a substantial impact on the deuterium fractionation in a dense core. As stated in \S \ref{sec:nn2d}, we expect the gas and dust to be well-coupled at the densities of Oph B2 such that $T_d = T_{gas}$, but specify in the following which of the gas or dust temperature is important in the following mechanisms. }First, at $T_{gas} \gtrsim 20$\,K, the \hd\, reaction in Equation \ref{eqn:hd} can proceed both forwards and backwards, resulting in no net increase in \hd, and consequently other deuterated species formed via reactions with \hd. Second, the CO which was deposited (adsorbed) as an icy mantle onto dust grains while at low temperatures will evaporate (desorb) back into the gas phase if the dust grains are heated, providing a destruction mechanism for \hd\, and again interrupting the deuteration chain. 


The CO depletion factor, $f_D (\mbox{CO})$, is defined as the canonical CO abundance, $X_{can}(\mbox{CO})$, divided by the CO abundance, $X(\mbox{CO}) = N(\mbox{CO})/N(\mbox{H$_2$})$, in the observed region:

\begin{equation}
f_D (\mbox{CO}) = \frac{X_{can}(\mbox{CO})}{X(\mbox{CO})}
\end{equation}

\noindent such that $f_D$ values are low where observed CO abundances are high. Low CO depletion factors can dramatically lower the \hd\, abundance at gas temperatures $T < 15$\,K \citep{caselli08}. Currently no estimates of $f_D$ exist in B2 due to a lack of data at spatial resolutions matching the available continuum data. The CO observations of the JCMT GBLS should enable the first calculations of $f_D$ at complementary spatial resolution, and allow further analysis of the relationship between CO and the deuterium distribution in B2. The extended distribution of \hd\, and \ddia\, across Oph B2 implies, however, that extensive depletion of CO has occurred. At densities of $n = 10^6$\,\cc, CO adsorption and desorption occur equally frequently when $T_d = 18$\,K \citep{visser09}. At $T_d > 18$\,K, CO will thus again become significantly more abundant in the gas phase, while at $T_{gas} > 20$\,K, enhanced deuterium fractionation via Equation \ref{eqn:hd} will cease to operate. 

An obvious potential source of heating is the presence of embedded protostars within Oph B2. In a 1D dynamic/chemical model of an isolated dense core which collapses to form a protostar, \citet[][see their Figure 2]{aikawa08} find temperatures greater than 18\,K can be found at radii $\sim 5000$\,AU ($\sim 0.02$\,pc) from the protostar at $\sim 10^5$\,yr after its formation. As a function of total luminosity of the core, $L_{core}$, however, their model predicts a CO sublimation radius of $\lesssim 1000$\,AU (0.005\,pc) for $L_{core} \sim 0.5 - 1\,L_\odot$. In a more specific example, \citet{stamatellos07} model the heating effect of an embedded Class I protostar ($L_{bol} = 10\,L_\odot$) on the surrounding gas in the Oph A Core, and find only a small temperature increase ($\lesssim 5$\,K), limited to the immediate vicinity of the protostar, in agreement with \citet{aikawa08}. 

Both YSOs in central B2 have been classified as embedded \citep{jorgensen08}, Class I sources \citep{enoch07}. The estimated bolometric luminosities, $L_{bol} \sim 0.5\,L_\odot$ and $\sim 1\,L_\odot$ for E32 and E33, respectively \citep{enoch07, vankempen09}, are much smaller than used in the heating model by \citet{stamatellos07}. The bolometric temperatures, $T_{bol} \sim 300$\,K and $\sim 500$\,K, suggest both are in the later stages of the Class I phase, as do their negative infrared spectral indices ($\alpha_{IR} = -0.03$ and $-0.12$). Sources with $-0.3 < \alpha_{IR} < 0$ are sometimes classified as `flat-spectrum' sources, and are thought to be an intermediate stage between Class I and Class II.  \citet{crapsi08} suggest that most or all flat-spectrum sources could be Class II objects seen edge-on. Based on these results, E32 and E33 are not likely to be deeply embedded within Oph B, and have ages $\sim 5 \times 10^5$\,yr or greater based on lifetime estimates of the different protostellar stages \citep{evans09}.

The low luminosities of E32 and E33 suggest that they should not have a large impact on the temperature of B2, in agreement with our temperature results in \one\, which showed very little variation in the gas temperature over Oph B. Figure \ref{fig:protall} shows that we find no trend in the kinetic temperature $T_K$, determined using \amm\, inversion transitions in \one, with protostellar distance above the $\sim 1-2$\,K uncertainties in the data. The derived temperatures are near the required values, however, and the  \amm\, temperatures may not be indicative of the conditions of the denser gas where we expect most of the deuterated species to exist. Depletion of \hd\, and \ddia\, via CO sublimation would thus require an increasing temperature gradient towards higher densities near the protostars, which is not unreasonable. 

Since CO also destroys \dia, significant evaporation of CO from dust grains should also be observable in the \dia\, distribution. Accordingly, in Figure \ref{fig:n2d}b we see a decrease in the integrated \dia\, 1-0 intensity towards E33. No good fit to the \dia\, spectra was found at this location, however, due to complicated line structures, such that we are not able to determine the \dia\, column density or abundance. Figure \ref{fig:protall}d shows no trend on larger scales of $X(\mbox{\dia})$ with protostellar proximity. The \dia\, 4-3 distribution shows \dia\, remains in the gas towards E33, but an analysis of the excitation of the line would be needed to determine the \dia\, column density from these data.

In summary, heating by an embedded protostar can produce slightly greater gas temperatures and evaporate CO from the surface of dust grains, which would decrease the deuterium fractionation in the gas. \showrev{Since we do not see evidence of a decrease in $X(\mbox{\dia})$ towards the protostars, the evaporation of CO from dust grains is likely not the major mechanism for the decrease in abundance of deuterated species. While we also don't see evidence of increased gas temperatures near the protostars in the \amm\, data, these temperatures reflect average values along the line-of-sight, and do not rule out increased temperatures at higher densities near the protostars. Given the strong dependence of increased deuterium fractionation on temperature, we expect this mechanism is likely very important in B2.}



\subsubsection{The Oph B2 outflow}

The deuterium fractionation may be impacted by the known, extensive CO outflow in B2 described in \S\ref{sec:linewidths}. It appears that the dense gas in eastern Oph B2 is affected by the outflow, given the larger line widths seen towards the protostars in high density tracers, changes in $v_{LSR}$, and the blue shoulder emission which is coincident with the eastern lobe of the outflow \citep{kamazaki03} (although large line widths could also indicate the presence of infall motions). We do not find evidence of outflow interaction with the dense gas in northwest Oph B2, suggesting that the outflow may be in front of or behind the bulk of the core gas. Very localized heating of the surrounding gas by the protostar could evaporate CO from the dust grains. The outflow, then containing entrained CO, could mix the CO with the dense core gas, destroying the deuterated species without an increase in the gas temperature on larger scales. Alternatively, shocks associated with the outflow could raise gas temperatures and reintroduce enough CO to the gas phase to affect the deuterium fractionation. Further analysis of the outflow, including observations of shock tracers, is needed to determine whether this scenario is reasonable, and can examine the impact of low-mass protostellar outflows on dense gas chemistry. \showrev{As discussed above, however, we do not find evidence for variation in the \dia\, abundance as would be expected if destruction due to gas-phase CO is the dominant mechanism for decreasing the abundance of deuterated species relative to their non-deuterated counterparts.}


\subsubsection{Ionization fraction and electron abundance $x_e$}

Variations in the ionization fraction in the dense gas can also affect the deuterium fractionation, in particular in regions of high CO depletion.  The electron abundance, $x_e = n_e/n_{\mbox{\tiny H$_2$}}$, is the ratio of the electron number density ($n_e$) with the number density of molecular hydrogen, $n_{\mbox{\tiny H$_2$}}$, and is equivalent to the ionization fraction. The deuterium fraction can be reduced by an increase in $x_e$, which in turn increases the rate of dissociative recombination \citep{caselli08}. The abundances of deuterated molecules can thus be used as a probe of $x_e$ in dense cores.  At high extinction towards starless cores, the creation of ions is expected to be dominated by the cosmic ray flux, with $x_e \propto n_{\mbox{\tiny H$_2$}}^{-1/2}$ \citep{mckee89}. 


Studies investigating $x_e$ in dense cores using the DCO$^+$/HCO$^+$ ratio have found average line-of-sight $x_e \sim 10^{-7}$ \citep[][and references therein]{bergin07}. In regions of high depletion, the column density ratio of \ddia\, with \dia\, is expected to be a better probe of $x_e$ than the ratio of DCO$^+$ with HCO$^+$. Since H$_3^+$ is expected to be the major molecular ion in cores strongly depleted in heavy elements, we can use observations of \hd, in conjunction with $R_D$, to set limits on $x_e$ assuming the numbers of positive and negative charges are approximately equal (i.e., the core gas is neutral). Given the abundances of \dia, \ddia and \hd\, in B2, we can thus derive a lower limit on $x_e$ following \citet{miett09}: 

\begin{equation}
x_e > X(\mbox{\dia}) + X(\mbox{\ddia}) + X(\mbox{H$_3^+$}) + X(\mbox{\hd})
\label{eqn:xe}
\end{equation}

\noindent where we neglect multiply deuterated forms of H$_3^+$. As discussed in \S\ref{sec:nh2d}, predictions of the o/p-\hd\, ratio vary from $\sim 0.1$ to $\gtrsim 1$, and we use a moderate o/p-\hd\, =  0.5 to estimate $X(\mbox{\hd})$ from $X(\mbox{ortho-\hd})$. In a simple, steady state analytical model, \citet{crapsi04} relate $R_D$ (where $R_D = N(\mbox{\ddia})\,/\,N(\mbox{\dia})$) to the relative \hd-H$_3^+$ abundance, $r = [\mbox{\hd}]/[\mbox{H$_3^+$}]$, and also neglect multiply deuterated forms of H$_3^+$ such that $R_D \approx r/(3 + 2r)$.  In an updated result, \citet{miett09} find $R_D \approx (r + 2r^2)/(3 + 2r + r^2)$. Using the \citeauthor{miett09} relation we then find a mean $r = 0.07$ across Oph B2, and a lower limit $x_e > 1.5 \times 10^{-8}$ through Equation \ref{eqn:xe}. We find small variations in the lower limit across B2, with a larger limit $x_e > 2.7 \times 10^{-8}$ towards Elias 32 and higher values in general near the 850\,\micron\, continuum peak. Detailed modeling, in conjunction with a measurement of the CO depletion factor in B2, is needed to improve this analysis of $x_e$ beyond a lower limit only.


\showrev{Low mass protostars emit x-rays, which could potentially increase the local ionization fraction and consequently decrease locally the deuterium fractionation of gas in Oph B2. In their study of the x-ray emission from young protostars in the central Ophiuchus region with ROSAT, \citet{casanova95} detected both Oph B2 protostars in the $1-2.4$\,keV range, finding luminosities (adjusted for extinction) $L_x = 30.7$\,erg\,s$^{-1}$ and $L_x = 29.2$\,erg\,s$^{-1}$ for E32 and E33, respectively. We can then estimate the x-ray ionization rate due to the protostars, $\zeta_x$, at a distance $r$  following \citet[][Equation 2]{silk83}. From Figure \ref{fig:protall}, we find trends in $R_D$ and $X(\mbox{ortho-\hd})$ to projected protostellar distances of order a few $\times 0.01$\,pc.  Assuming a constant, uniform density $n = 10^5$\,\cc, at $r = 0.02$\,pc we find $\zeta_x \sim 9 \times 10^{-17}$\,s$^{-1}$, and at $r = 0.04$\,pc we find $\zeta_x \sim 8 \times 10^{-18}$\,s$^{-1}$. The fractional ionization based on these rates, again following \citeauthor{silk83}, is $x_e \sim 9 \times 10^{-8}$ and $x_e \sim 3 \times 10^{-8}$ at $r = 0.02$\,pc and $r = 0.04$\,pc, consistent with our lower limits calculated above. Clumpy, rather than uniform gas could allow x-rays to penetrate further into the core, potentially increasing the fractional ionization to greater distances. The standard ionization rate due to cosmic rays alone $\zeta_{CR} = 1.3 \times 10^{-17}$\,s$^{-1}$ \citep{herbst73}, leading to a fractional ionization $x_e \sim 4 \times 10^{-8}$ for $n = 10^5$\,\cc\, \citep{mckee89}. Our simple calculation shows that at distances comparable to where we find variation in the deuterium fractionation in Oph B2, the protostellar x-ray ionization rate and resulting fractional ionization may be comparable to or greater than expected due to cosmic rays alone. This result suggests that x-ray ionization is a viable mechanism to depress the deuterium fractionation close to embedded protostars. Further work is needed to directly measure the fractional ionization of dense gas near protostars, and to quantify the effect of a greater ionization rate on the chemical network of deuterated species in high density gas. }

\showrev{\subsubsection{Other possibilities}}

\showrev{While Figure \ref{fig:protall} suggests that the distribution of enhanced deuterium fractionation of H$_3^+$ and \dia\, in Oph B2 is tied to the presence of the Class I protostars associated with the core, other factors may influence the deuteration chemistry in this complex environment. For completeness, we briefly mention two other possibilities. First, line-of-sight inhomogeneities in Oph B2 would impact our estimate of the deuterium fractionation, as our analysis assumes constant excitation and abundance conditions along the line of sight. Detailed physical and chemical modeling may be useful in probing the effects of possible inhomogeneities, but is beyond the scope of this paper due to the complexity of Oph B2. Second, differences in the evolutionary timescale may be important in Oph B2, and in other multiple-star-forming objects. While several protostars have already formed out of the core gas (particularly in the south), there remain multiple dense, starless clumps within B2 which may yet form stars. The timescale on which different parts of the core condense and collapse may impact the dense gas chemistry and produce variations in the deuterium fractionation across Oph B2. }

\showrev{In summary, there a number of different mechanisms by which the deuterium fractionation of \dia\, and H$_3^+$ may be decreased near young protostars. We feel that a local increase in the gas temperature is the most likely explanation of this trend, but our simple calculation of the expected increase in the ionization fraction above the standard cosmic ray value is intriguing and worth further study. }

\section{Summary}
\label{sec:summary}

In this paper we have presented the results of \ddia\, 3-2, \dia\, 4-3 and ortho-\hd\, $1_{11} - 1_{10}$ mapping of the cluster-forming Ophiuchus B Core. In general, our results show that care must be taken when using deuterated species as a probe of the physical conditions of dense gas conditions in star-forming regions. We summarize in detail our results below. 

1. Significant emission from the deuterated species \ddia\, and \hd\, extends over several square arcminutes in B2. In particular, the extent of the \hd\, emission is the largest yet mapped. Both species' integrated intensity distributions are offset from the continuum emission, and avoid entirely parts of B2 near embedded protostars and the peak of continuum emission from cold dust. 

2. Through fitting of the spectral lines, we find greater \ddia\, line widths near protostars embedded in B2, suggestive of protostellar influence on the dense gas. Complicated line structures are found, including blue line shoulders, in all three tracers. We estimate the non-thermal line widths for the species observed here and compare with \amm\, and \dia\, results from \one\, and \two.  In B2, non-thermal line widths decrease with increasing critical density of the molecular tracer, but remain transonic or only slightly subsonic even at high densities. This result is in contrast with the nearly thermal line widths found in many isolated cores. 

3. We test a method to estimate the gas temperature by equating the non-thermal line widths of \ddia\, and \hd\, and solving for $T_K$. We find both realistic ($T_K \sim 11$\,K) and unrealistic ($T_K < 0$\,K and $T_K \gtrsim 25$\,K) values in B2. Complicated line structure in B2 likely affects this analysis, but we expect good results can be found in regions with simple velocity structure. 

4. The deuterium fraction of \dia\, $R_D = N(\mbox{\ddia})/N(\mbox{\dia}) \sim 0.03$ on average where \ddia\, is detected. Small-scale features of enhanced $R_D$ are seen in B2, but most do not correlate with previously identified continuum clumps or \dia\, clumps. This $R_D$ is consistent with previous results in protostellar cores, and is also within the lower range of values found for starless cores.

5. The average ortho-\hd\, abundance, $X(\mbox{ortho-\hd}) \sim 3 \times 10^{-10}$, given an assumed excitation temperature of 12\,K. An accurate measure of the ortho- to para-\hd\, ratio is needed to convert to total \hd\, abundance, and has been found in previous studies to vary between 0.1 and $> 1$. Future observations of para-\hd\, transitions, if possible, are needed to constrain this value in dense cores.

6. An anti-correlation is found with $R_D$ and $X(\mbox{ortho-\hd})$ and projected distance to the nearest embedded protostar in B2. This relation indicates that the embedded protostars in B2 are affecting the internal chemistry. Possible mechanisms are through gas heating, evaporation of CO off dust grains, CO mixing caused by the B2 outflow, or by increasing the local ionization fraction. Gas temperatures determined through \amm\, analysis in \one\, are $< 18-20$\,K required to evaporate CO and reduce deuterium fractionation, but \amm\, may not be a good tracer of conditions in the high density gas in Oph B2. \showrev{We calculate that the x-ray flux of the embedded protostars in B2 could increase the fractional ionization of the core gas beyond that expected solely due to cosmic ray ionization to distances $\sim 0.02 - 0.03$\,pc, similar to the distances over which we find variation in the deuterium fractionation. We suggest that a temperature increase is the dominant parameter in this trend, but further research on the effect of young stars and their outflows on both the local ionization fraction and the chemistry of deuterated species in protostellar cores is needed. }

7. We use the observed $R_D$ and $X(\mbox{ortho-\hd})$ to estimate a lower limit on the ionization fraction in B2, and find $x(e) > 1.5 \times 10^{-8}$ given an assumed ortho- to para-\hd\, ratio of 0.5. 

\acknowledgments

We thank the anonymous referee and E. Feigelson for comments which improved the paper. We also thank H. Kirk for providing SCUBA maps of the regions observed and H. Weisemeyer for help with the IRAM observations. The National Radio Astronomy Observatory is a facility of the National Science Foundation operated under cooperative agreement by Associated Universities, Inc. The James Clerk Maxwell Telescope is operated by the Joint Astronomy Centre on behalf of the Particle Physics and Astronomy Research Council of the United Kingdom, the Netherlands Association for Scientific Research, and the National Research Council of Canada. RKF acknowledges financial support from the University of Victoria and the National Research Council Canada Graduate Student Scholarship Supplement Program. We also acknowledge the support of the National Science and Engineering Research Council of Canada. This research is supported in part by the National Science Foundation under grant number 0708158 (TLB).

{\it Facilities:} \facility{JCMT}, \facility{IRAM}

\bibliographystyle{apj}
\bibliography{biblio}

\begin{deluxetable}{lcccccc}
\tablecolumns{7}
\tablewidth{0pt}
\tablecaption{Observed Species, Transitions and Rest Frequencies \label{tab:obs}}
\tablehead{
\colhead{Species} 	& \colhead{Transition} 	& \colhead{$F_1^\prime$} & \colhead{$F^\prime$} & \colhead{$F_1$} & \colhead {$F$} & \colhead{Frequency} \\
\colhead{}			& \colhead{}			&                                              &                                        &                                &                            & \colhead{GHz}}
\startdata
\ddia				& $3-2$				& 4 & 5 & 3 & 4 & 231.321906 \tablenotemark{a} \\
\dia				& $4-3$				& 5 & 5 & 4 & 4 & 372.672494 \tablenotemark{b}	\\
\hd				& $1_{11} - 1_{10}$		&    &    &     &    & 372.421385 \tablenotemark{c} \\
\enddata
\tablenotetext{a}{\citet{gerin01}}
\tablenotetext{b}{\citet{pagani09}}
\tablenotetext{c}{\citet{amano05}}
\end{deluxetable}

\begin{deluxetable}{ll|cccc|cccc|c}
\tablecolumns{11}
\tablewidth{0pt}
\tablecaption{Mean \amm, \dia, \ddia\, and \hd\, $v_{LSR}$ and $\Delta v$ in Oph B2 \label{tab:tracers}}
\tablehead{
\colhead{Species} & \colhead{Transition} &
				\colhead{$v_{LSR}$} & \colhead{} & \colhead{} & \colhead{} & 
				\colhead{$\Delta v$}   & \colhead{} & \colhead{} & \colhead{} &
				\colhead{$n_{cr}$} \\ 
\colhead{} & \colhead{} 	& \colhead{\kms} & \colhead{} & \colhead{} & \colhead{} &
					   \colhead{\kms} & \colhead{} & \colhead{} & \colhead{} & \colhead{\cc} \\
\colhead{} & \colhead{} &
				\colhead{Mean} & \colhead{RMS} & \colhead{Min} & \colhead{Max} &
			         \colhead{Mean} & \colhead{RMS} & \colhead{Min} & \colhead{Max} &
			         	\colhead{}}
\startdata
\amm\, & $(J,K) = (1,1)$ & 4.14 & 0.19 & 3.85 & 4.58 & 0.87 & 0.16 & 0.42 & 1.11 & $2 \times 10^4$\,\tablenotemark{a} \\
\dia\, & $J=1-0$       	& 4.04 & 0.20 & 3.61 & 4.43 & 0.71 & 0.22 & 0.38 & 1.42 & $1.4 \times 10^5$\,\tablenotemark{b} \\
\hd\, & $J_{K_{-1}K_1}$ & & & & & & & & & \\
 & = $1_{11} - 1_{10}$ 
		    	 	& 4.01 & 0.24 & 3.65 & 4.50 & 0.74 & 0.19 & 0.51 & 1.36 & $\sim 10^5$\,\tablenotemark{c} \\
\ddia\, & $J=3-2$     	& 4.02 & 0.19 & 3.67 & 4.44 & 0.53 & 0.20 & 0.25 & 1.36 & $8 \times 10^5$\,\tablenotemark{d} \\
\dia\, & $J=4-3$\tablenotemark{e}       	& 4.05 & 0.22 & 3.39 & 4.43 & 0.69 & 0.18 & 0.47 & 1.14 & $7.7 \times 10^6$\,\tablenotemark{b} \\
\enddata
\tablecomments{The velocity resolution of the \amm\, observations was 0.3\,\kms. All other species were observed with 0.1\,\kms\, velocity resolution or better.}
\tablenotetext{a}{\citet{caselli02}}
\tablenotetext{b}{\citet[$n_{cr}$ calculated using Einstein A coefficients and collision rates $\gamma_{ul}$ at 10\,K from the Leiden Atomic and Molecular Database; ][]{schoier05}}
\tablenotetext{c}{\citet{caselli08}}
\tablenotetext{d}{Calculated using Einstein A coefficients and collision rates for \dia\, 3-2 at 10\,K \citep{schoier05}. \citet{daniel07} (see their Figure 17) found that despite small differences in Einstein A coefficients and collision rates, critical densities of \dia\, and \ddia\, are similar for transitions other than $J = 1 - 0$.}
\tablenotetext{e}{Values calculated from a Gaussian fit due to a lack of visible hyperfine structure. The reported line widths may be greater by $\sim 10-25$\,\%\, than the true values; see \S\ref{sec:line_fit}.}
\end{deluxetable}

\begin{deluxetable}{cccc}
\tablecolumns{4}
\tablewidth{0pt}
\tablecaption{Impact of $T_{ex}$ on $\tau$, $Q_{rot}$ and $N(\mbox{ortho-\hd})$ \label{tab:tex_test}}
\tablehead{
\colhead{$T_{ex}$} & \colhead{$\tau$} & \colhead{$Q_{rot}$} & 
	\colhead{$N(\mbox{ortho-\hd})$} \\
\colhead{K} & \colhead{ } & \colhead{ } & \colhead{$10^{13}$\,cm$^{-2}$} }
\startdata
   7  & 0.41   &   9.7 & 1.60 \\
   8  & 0.27   & 10.0 & 1.11 \\
   9  & 0.20   & 10.2 & 0.86 \\
  10 & 0.15   & 10.5 & 0.71 \\
  11 & 0.12   & 10.8 & 0.61 \\
  12 & 0.10   & 11.0 & 0.54 \\
  13 & 0.09   & 11.3 & 0.48 \\
  14 & 0.08   & 11.5 & 0.45 \\
  15 & 0.07   & 11.7 & 0.42 \\
\enddata
\tablecomments{Values were calculated for a line strength $T_{MB} = 0.5$\,K and width $\Delta v = 0.7$\,\kms.}
\end{deluxetable}

\

\end{document}